\documentclass[10pt,journal,compsoc]{IEEEtran}
\ifCLASSOPTIONcompsoc
  \usepackage[nocompress]{cite}
\else
  \usepackage{cite}
\fi
\ifCLASSINFOpdf
\else
\fi
\usepackage{url}

\usepackage[plain]{algorithm}
\usepackage{breakurl}
\usepackage{subfig}
\usepackage{graphicx}
\usepackage{fancyhdr}
\usepackage{xspace}
\usepackage{pseudocode}
\usepackage{tabularx}
\usepackage{rotating}
\usepackage{multirow}
\usepackage{array}
\usepackage[usenames,dvipsnames,svgnames,table]{xcolor}
\usepackage{booktabs}
\usepackage{changepage}
\usepackage{amssymb,amsmath}
\usepackage{stfloats}
\usepackage{amsmath}
\usepackage{tablefootnote}
\usepackage{comment}
\usepackage[font={small,it}]{caption}
\usepackage{textcomp}
\usepackage{microtype}
\usepackage{wrapfig,lipsum,booktabs}
\usepackage{longtable}
\usepackage{caption}
\usepackage{times}
\usepackage[T1]{fontenc}
\usepackage{hyperref}
\usepackage{epstopdf}

\newcommand{\prover}{$\mathcal P$\xspace}
\newcommand{\verifier}{$\mathcal V$\xspace}
\newcommand{\comparator}{$\mathcal C$\xspace}

\newcommand{\zeromodal}{zero-modality\xspace}
\newcommand{\singlemodal}{single-modality\xspace}
\newcommand{\SingleModal}{Single-Modality\xspace}
\newcommand{\multimodal}{multi-modality\xspace}

\newcommand{\altitude}{{\sf Al}\xspace}
\newcommand{\humidity}{{\sf H}\xspace}
\newcommand{\gas}{{\sf G}\xspace}
\newcommand{\temperature}{{\sf T}\xspace}
\newcommand{\audio}{{\sf Au}\xspace}
\newcommand{\bluetooth}{{\sf B}\xspace}
\newcommand{\wifi}{{\sf W}\xspace}

\newif\ifrevision
\revisionfalse

\ifrevision
	\newcommand{\verOne}[1]{{#1}}
	\newcommand{\verTwo}[1]{{\color{blue}#1}}
\else
	\newcommand{\verOne}[1]{#1}	
	\newcommand{\verTwo}[1]{#1}
\fi

\makeatletter

\def\ps@IEEEtitlepagestyle{%
  \def\@oddfoot{\mycopyrightnotice}%
  \def\@evenfoot{}%
}
\def\mycopyrightnotice{%
  {\footnotesize \textit{In Proceedings of IEEE Transactions on Mobile Computing, IEEE TMC 2018. \hfill}}
  \gdef\mycopyrightnotice{}
}

\begin{document}
%

\title{\verOne{Sensor-based Proximity Detection in the Face of Active Adversaries}}

\author{
	Babins~Shrestha,
                          Nitesh~Saxena,~\IEEEmembership{Member,~IEEE},
                          Hien~Thi~Thu~Truong, 
                          N.~Asokan,~\IEEEmembership{Fellow,~IEEE} 
\IEEEcompsocitemizethanks{\IEEEcompsocthanksitem Babins Shrestha is with VISA Inc., USA. Email: babishre@visa.com%
\IEEEcompsocthanksitem Nitesh Saxena is with University of Alabama at Birmingham, Birmingham, AL 35294. 
E-mail: saxena@uab.edu
\IEEEcompsocthanksitem Hien Thi Thu Truong is with NEC Laboratories Europe at Heidelberg, Germany. 
E-mail: hien.truong@neclab.eu%
\IEEEcompsocthanksitem N. Asokan is with Aalto University, Finland. 
E-mail: asokan@acm.org}
\thanks{}}

%

\IEEEtitleabstractindextext{%
\begin{abstract}

Context-centric \verOne{sensor-based} proximity detection (or, \verOne{contextual} co-presence detection) is a promising
approach to defend against \textit{relay attacks} in many mobile authentication
systems, especially against \textit{unattended terminals} (such as cars parked in unmonitored parking lots, 
remote gas station pumps, or stolen laptops).  
Prior work demonstrated the effectiveness of a variety of contextual sensor
modalities for this purpose, including audio-radio environment
(ambient audio, Wi-Fi, Bluetooth and GPS, and combinations thereof) and
physical environment (temperature, humidity, gas and altitude, and
combinations thereof).  
%

In this paper, we present a systematic assessment of such co-presence detection in the presence of
a strong, context-manipulating attacker against unattended terminals. \textit{First}, we show that it is feasible to
\textit{manipulate}, \textit{consistently control} and \textit{stabilize} the
readings of different acoustic and physical environment sensors (and even
multiple sensors simultaneously) using low-cost, off-the-shelf equipment. 
Specifically, 
we show that it is possible to control the temperature using a home-grade hair
dryer, affect the gas readings using a smoking cigarette, impact
the altitude/pressure with a simple air compressor, or relay audio signals recorded at one end to the other
thereby causing both sides to perceive a very similar acoustic environment.
\textit{Second}, based on these capabilities and the strengthened threat model, we 
show that an attacker who can manipulate the context gains a significant
advantage in defeating \verOne{contextual} co-presence detection. For systems that
use multiple sensors, we investigate two sensor fusion approaches based on
machine learning \verOne{classification} techniques -- \textit{features-fusion} and
\textit{decisions-fusion}, and show that both are vulnerable to \verOne{context manipulation}
attacks but the latter approach can be more resistant in some cases.
We further consider other defensive approaches that may be used 
to reduce the impact of even such a strong context-manipulating attacker. 

Our work represents the first concrete step towards analyzing, extending and
systematizing prior work on contextual co-presence detection under a stronger,
but realistic adversarial model.  


\end{abstract}

}

\maketitle
\IEEEdisplaynotcompsoctitleabstractindextext

\IEEEpeerreviewmaketitle

\section{Introduction}
\label{sec:intro}
Authentication is critical to many mobile and wireless systems where one
communicating device (prover \prover) needs to validate its identity to the
other (verifier \verifier).  Traditional cryptographic authentication typically
involves a challenge-response protocol whereby \prover proves the possession of
the key $K$ that it pre-shares with \verifier by constructing a valid response
to a random challenge sent by \verifier.  
Examples of systems where such
authentication is deployed include payment transactions between NFC/RFID
devices and point-of-sale systems, and zero-interaction authentication
\cite{CN02} scenarios between a token and a terminal (e.g., phone-laptop, or key-car).  Unfortunately, the security and usability benefits
provided by these authentication systems can be subverted by means of \textit{relay attacks}, as demonstrated by prior research (e.g.,
\cite{drimer2007distance,francillon2010carkey}), which involve two non
co-present colluding attackers, one near \prover and one near \verifier, simply
relaying protocol messages back and forth between \prover and \verifier.
A known defense to relay attacks is
\emph{distance bounding}, where a challenge-response
authentication protocol allows \verifier to measure an upper-bound of its
distance from \prover \cite{brands1993distance}.  Using this protocol,
\verifier can verify whether \prover is within a close proximity thereby
detecting the presence of relay attacks
\cite{drimer2007distance,francillon2010carkey}.  \verOne{Although distance bounding
systems are gradually becoming commercially available \cite{3db}}, they may not be feasible on all \verOne{cost-sensitive} commodity devices (such as
smartphones or payment tokens) due to their
sensitivity to measurement errors (of elapsed time). 

The presence of ubiquitous and low-cost sensing capabilities on many
modern mobile devices has facilitated a potentially more viable relay attack defense
\cite{DBLP:conf/esorics/HaleviMSX12,varshavsky07amigo,Narayanan11,FrancisHMM-2010-rfidsec}.
This defense leverages the notion of ``context'' derived from on-board device
sensors based on which \prover-\verifier proximity, or lack of it, could be
determined. In other words, in a benign setting, where \prover and \verifier
are co-present, both would record a similar context with a high probability.
In contrast, if the system is subject to a relay attack, and \prover and
\verifier are non co-present, devices' context should be different with a high
probability. 

Extensive recent prior work demonstrated the feasibility of using different
types of sensor modalities for such \textit{contextual co-presence detection},
including audio\footnote{The work presented in \cite{SP} also makes use 
of audio to detect whether an authentication token is present near the browser, forming 
a two-factor authentication scheme. However, the focus of this work 
is not to defeat proximity or relay attacks but rather to defeat remotely located adversaries, and it is therefore
not studied in this paper.}
 \cite{DBLP:conf/esorics/HaleviMSX12}, radio (\textit{Wi-Fi}
\cite{varshavsky07amigo}, \textit{Bluetooth} \cite{TruongPerCom14} and
\textit{GPS} \cite{FrancisHMM-2010-rfidsec}, and the physical environment
(\textit{temperature}, \textit{humidity}, \textit{gas} and
\textit{altitude/pressure}) \cite{ShresthaFC2014}.  Many single modalities,
such as audio and Wi-Fi, were shown to be performing quite well for contextual
co-presence detection resulting in low \textit{false negatives} (i.e.,
rejecting a co-presence instance; a measure of usability) and low \textit{false
positives} (accepting a non co-presence instance; a measure of security). In
addition, \textit{fusion} of multiple modalities, including combination of
audio-radio \cite{TruongPerCom14}, and combination of physical sensors
\cite{ShresthaFC2014}, has been shown to further reduce false negatives and
false positives.


\verOne{\smallskip \noindent {\textsc{Our Work vs. Related Work:}}} The focus of prior work cited above on contextual co-presence detection largely centered on
evaluating the system's security under the assumption that it is very hard to
manipulate the contextual environment (i.e., it considered only a Dolev-Yao
attacker \cite{Dolev:1981:SPK:891726}). 
%
In this paper, we are extending this model to the realm of a
context-manipulating attacker, especially against unattended terminals.
Vehicles parked in underground parking lots/decks represent an apt
example of unattended verifiers\footnote{Clarke \cite{clarke2002thefts} 
reports that most of the theft happens at unattended places such as 
parking lots where there is rarely much surveillance.}.  
Relay attacks against such vehicles have
already been demonstrated in the literature \cite{francillon2010carkey} and reportedly being
executed in the wild by the car criminals \cite{car-relay}. Other examples
include stolen laptops in a zero-interaction authentication system.  Payment
scenarios, such as those involving parking meters or remote gas station pumps,
also involve unattended payment terminals and are thus also subject to our
study.

\smallskip \noindent {\textsc{Our Contributions:}} 
\verOne{The main focus of our work is on the assessment of existing contextual co-presence
detection systems against active attackers.}
The primary contributions of this paper are two-fold:


\smallskip \noindent \textbf{1.\ {Building Simple Context Manipulation
Attacks against Unattended Terminals}}: We show that it is feasible to manipulate the readings of different
sensors (and combinations thereof) using
low-cost, off-the-shelf equipment, representing a 
realistic attacker against unattended terminals. We demonstrate attacks against a variety of modalities
studied in prior work including audio, radio (Bluetooth/Wi-Fi), and physical
(temperature, humidity, gas and altitude).  

In particular, 
we demonstrate how an attacker in
close proximity of the sensors can successfully control, manipulate and stabilize the
physical environment ``seen'' by these sensors, without the need to manipulate
the global surrounding environment or compromise the devices/sensors
themselves.
For instance, we show
that it is possible to control the temperature using a home-grade hair dryer
or ice cubes, affect the gas readings using a smoking cigarette, or impact the
altitude/pressure with a simple air compressor made up of a plastic bag. 
\verOne{We also observe that with practice, an attacker can increase his effectiveness 
surprisingly quickly, suggesting that our attacks should be taken as a lower bound -- 
dedicated attackers are likely to fare significantly better than our results show.}
Our attacks are described in Section \ref{sec:manipulation_attack}.


\smallskip \noindent \textbf{2.\ {Quantifying the Security of Co-Presence Detection in the Presence of Context
Manipulations}}: Based on the above manipulation capabilities, we
comprehensively examine and \textit{quantify} the advantage a
\textit{\multimodal attacker} can have in defeating co-presence detection over a
\zeromodal attacker (one studied in prior work). 
\verOne{A \textit{\multimodal attacker} can manipulate multiple sensor
modalities simultaneously while a \zeromodal attacker cannot manipulate any modality.}
To accomplish this, we
re-orchestrated the co-presence detection approaches based on machine learning \verOne{classification} techniques in
audio-only \cite{DBLP:conf/esorics/HaleviMSX12}, audio-radio
\cite{TruongPerCom14}, physical \cite{ShresthaFC2014} and
(\textit{a newly-proposed}) audio-radio-physical systems, in a way that non
co-present data samples were manipulated for different modality combinations.
Our results show that the attacker advantage increases \textit{many-folds} in several
cases (Table \ref{tbl:all-mani} quantifies the attacker success rates). 

For systems that use multiple modalities, we investigate two different sensor
fusion approaches -- \textit{features-fusion} (proposed in
\cite{TruongPerCom14}) and \textit{decisions-fusion} based on majority voting,
and show that both approaches are vulnerable to contextual attacks but the
latter can be more resistant in some cases, at the cost of slight degradation
in usability.  Our detailed analysis is presented in Section
\ref{sec:analysis}.

\smallskip \noindent {\textsc{Broader Impact and Lessons Learned:}} Our work
represents the first concrete step towards analyzing, extending and
systematizing prior work on contextual co-presence detection under a stronger,
but realistic adversarial model.  
It suggests that tampering with context may
not be very difficult, and the security offered by
contextual co-presence detection therefore weakens. 

Although a sophisticated
attacker would likely fare better at manipulating the context (compared to our
attacks), we also suggest potential strategies (including {decisions-fusion}) 
that may still be used to strengthen the security of co-presence
detection against a \multimodal attacker (Section \ref{sec:discussion}). 
At a broader level, our work calls the security of contextual co-presence detection
into question, and motivates the need of re-evaluating the security of other
context-centric systems in the face of context manipulation. 
For instance, our work may be extended to analyze the security of other
promising context-based systems such as contextual access control
\cite{context-asiaccs2014} with respect to context-manipulating adversaries.



\section{Background and Models}

\label{sec:background}

\subsection{Relay Attacks \& Contextual Co-Presence Detection}
\label{subsec:CCP}

The goal of the adversary against a challenge-response authentication system
is to fool \verifier into
concluding that \prover is nearby and thus needs access to \verifier even when
\prover is actually far away.  The attacker possesses standard Dolev-Yao
capabilities~\cite{Dolev:1981:SPK:891726}: it has complete control of the
communication channel over which the authentication protocol between \prover
and \verifier is run but does not have physical possession of \prover nor is
able to compromise (e.g., through malware) either \prover or \verifier. 

The attacker could take the form of a
``ghost-and-leech''~\cite{Kfir:2005:PVP:1128018.1128470} duo ($A_p$, $A_v$)
such that $A_p$ (respectively $A_v$) is physically close to \prover
(\verifier), and $A_p$ and $A_v$ communicate over a high-speed connection.
Such an adversary pair can 
compromise the security of traditional challenge-response authentication by
simply initiating a protocol session between \prover and \verifier, relaying
messages between them, leading \verifier to conclude that \prover is in
proximity. This is an attack applicable to zero-interaction authentication
systems. A similar attack applies to proximity-based payment systems
\cite{mafia,drimer2007distance}.

\begin{figure*}[htpb] \centering
	\includegraphics[width=\textwidth]{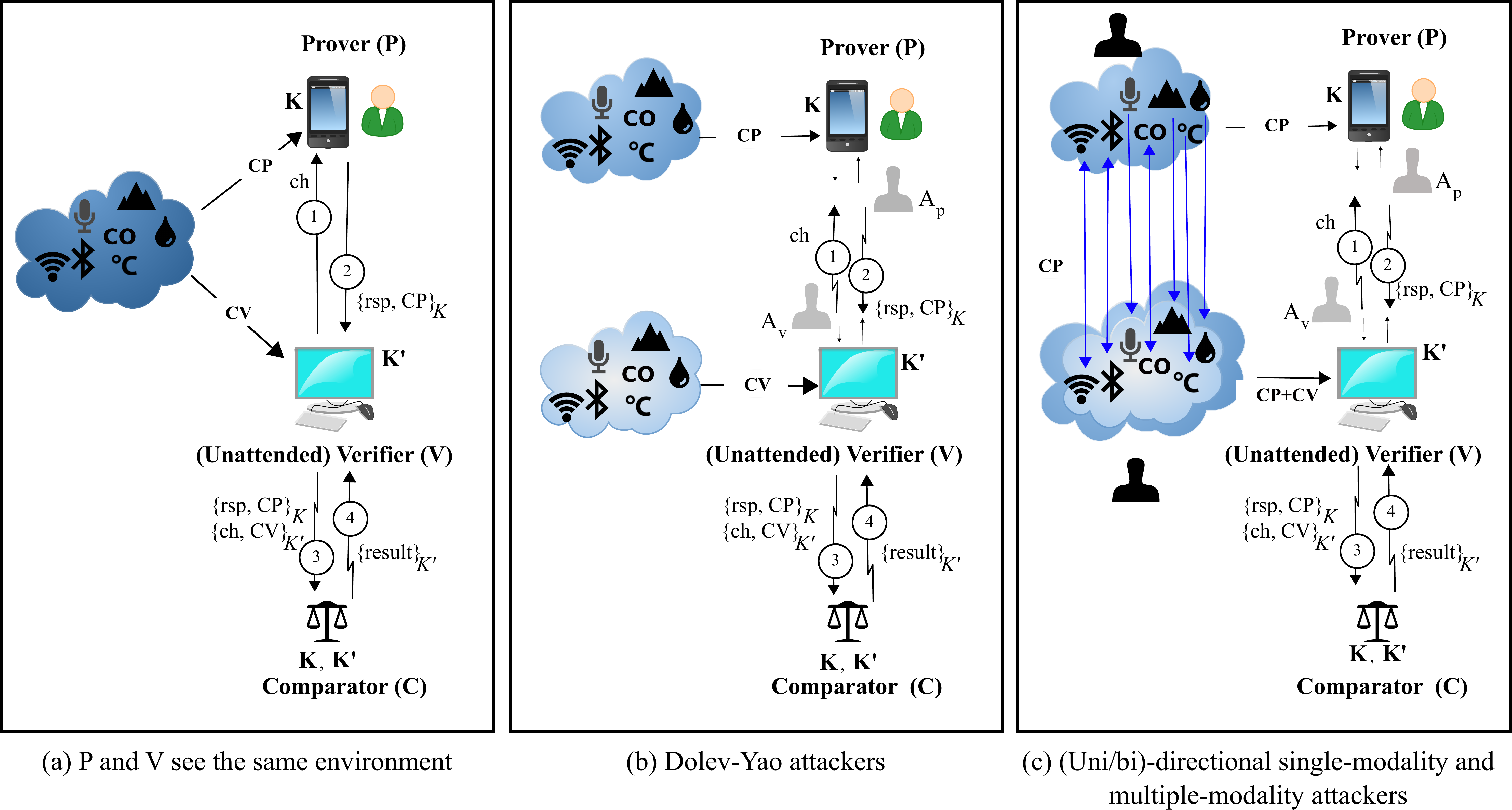} \vspace{-5mm}\caption{System
        model of proximity based authentication with contextual co-presence. In our work, the unidirectional single-modal attacker model 
\cite{TruongPerCom14} is extended to bidirectional and multiple-modality attackers (highlighted with blue arrows). Radio sensors and Gas are subject to bidirectional attacks in our new model.} \label{fig:attack-model}
\end{figure*}

Co-presence detection schemes aim to address such relay attacks.
Fig.~\ref{fig:attack-model}(a) shows a typical system model of an
authentication/authorization protocol using contextual co-presence, adapted
from \cite{TruongPerCom14}.  
In this defense, \prover (respectively \verifier) pre-shares a key
$K$ ($K'$) with a ``comparator'' \comparator (which may be part of \verifier or
a separate entity, depending on the scenario).  When \prover sends a trigger to
\verifier, it responds with a challenge $ch$. \prover and \verifier then
initiate context sensing for a fixed duration $t$. \prover computes a response
$rsp$ (using $K$), appends it to the sensed context information $CP$ and sends
both \verifier, protected by $K$. \verifier forwards this to \comparator.
In the meantime, \verifier finishes sensing its own context and sends the
resulting context data $CV$ protected using $K'$ to \comparator. 
\comparator then recovers $CP$, $CV$, $ch$ and $rsp$. It checks the
validity of $rsp$ and compares if $CP$ is sufficiently similar to $CV$.  If
both checks succeed, \comparator concludes that \prover and \verifier are
co-present.  When \comparator is integrated with \verifier, $K'$ is not used.
Fig.~\ref{fig:attack-model}(b) shows how contextual co-presence can thwart
a Dolev-Yao relay attacker.



Prior work has proposed the use of different sensor modalities for such
co-presence detection: ambient audio -- \audio
\cite{DBLP:conf/esorics/HaleviMSX12}, radio context including Wi-Fi -- \wifi and
Bluetooth -- \bluetooth \cite{TruongPerCom14}, and physical environmental attributes, temperature -- \temperature, humidity -- \humidity,
concentration of gases -- \gas and altitude -- \altitude
\cite{ShresthaFC2014}.

\subsection{Threat Model: \SingleModal Contextual Attacker}
\label{subsec:threatmodel} 

Our focus is on a context-manipulating attacker against co-presence detection (going beyond a Dolev-Yao attacker).
Truong et al. \cite{TruongPerCom14} briefly explored the problem of
characterizing such a contextual attacker. They only consider an attacker who
is capable of manipulating a single sensor modality at a time
(``\singlemodal attacker'', in our parlance). Again, in this model, an attacker cannot 
compromise \prover and \verifier devices.
Based on the rationale that \verifier is often \textit{unattended}, whereas \prover
is in the possession of a human user, they speculated that the context
attacker can manipulate context without detection only in one direction.
More precisely, they modeled a \singlemodal attacker as follows:
\begin{itemize}
\itemsep0em
\item $A_p$, $A_v$ can measure the context information that \prover, \verifier
  would sense, respectively. 
\item $A_v$ can fool \verifier into sensing the context information $A_v$
  chooses. Specifically $A_v$ can receive context information from
  $A_p$ and reproduce it near \verifier. 
\item $A_v$ ($A_p$) cannot suppress any contextual
  information from being sensed by \verifier (\prover).
\end{itemize}

Fig.~\ref{fig:attack-model}(c) illustrates this threat model.  
Later in
Section \ref{sec:analysis}, based on our context-manipulation attacks presented
in Section \ref{sec:manipulation_attack}, this current model will be extended,
to incorporate \multimodal attackers, who can perform the above (\singlemodal)
tasks corresponding to multiple modalities simultaneously. 

\section{Context Manipulation Attacks and \\Attack Experiments}
\label{sec:manipulation_attack}
In this section, we present our context manipulation attacks against audio,
radio and physical sensor modalities, and their various combinations.
\verOne{We explain the concepts underlying the attack approaches and present attack 
experiments when needed.
``Modality'' refers to the \emph{type} of sensor that generates the raw input data used for 
modeling the ambient context \cite{siegwart2011introduction}.}

\subsection{Manipulating Audio Sensor Modality}
\label{sec:audiomanipulate}
To manipulate ambient audio, an adversary must find a way to make ambient audio
on one side similar to that on the other side.  Recall from
Section~\ref{sec:background} that our threat model allows the attacker to add
to the ambient audio at \verifier's side without being noticed, allowing him to
relay/stream the ambient audio in real-time from \prover's side to \verifier's
side thereby causing the features used for audio correlation almost match at
both sides.  The assumption that manipulating audio at \verifier's side can
go undetected is valid since \verifier  may be unattended in many scenarios
(as our model in Section \ref{sec:background} assumed).  The attacker duo can
use any reliable audio streaming tool 
to stream the audio from \prover's side to \verifier's side.  They can execute
this attack conveniently using mobile phones and wireless data connection.
We evaluated how well such an attacker can succeed in fooling audio-based
co-presence detection by streaming ambient audio 
using Skype \cite{skype}.  We use the features and classifier described in prior work
\cite{DBLP:conf/esorics/HaleviMSX12}. Our results are presented in Section
\ref{sec:audio-only}.

\subsection{Manipulating Radio-Frequency Sensor Modalities}
\label{sec:radio}
Prior work suggests that
manipulating the radio context is possible in general.
The work presented in \cite{Tippenhauer2009WLanAttacks} describes attacks on a
public Wi-Fi based positioning system. They used a Linux laptop as an Access
Point (AP) with the Scapy packet manipulation program \cite{scapy} to spoof Wi-Fi APs.
Similarly, spoofing bluetooth device addresses has already been demonstrated in
prior work \cite{levi2004bluetoothRelayAttack,TruongPerCom14}, both of which
reported bluetooth-based relay attacks. An attacker can control the received
signal strength by controlling the transmission power of his masquerading
devices.  Therefore, we conclude that the threat model assumed in
\cite{TruongPerCom14} (see Section~\ref{subsec:threatmodel}) is reasonable.
Furthermore, in the case of Radio Frequency (RF) sensor modalities, it is reasonable to assume
that an attacker can also manipulate the RF environment at \prover's end
without being noticed (since radio waves are imperceptible to human users).
Therefore, limiting the attacker to unidirectional manipulation only is too
restrictive. 

We tested the feasibility of Wi-Fi spoofing ourselves, and studied how it can be
used to match the Wi-Fi context at two ends.  In our experiment, we used a
Linksys router (WRT54G) to create a spoofed hotspot.  We flashed DD-WRT
firmware \cite{ddwrt} to the router since the default firmware did not allow us to
spoof the Basic Service Set Identifier (BSSID).
The router used in our experiment is portable, easily available in the market, and much
cheaper than other devices which can also be used to spoof the
hotspot such as laptops or smartphones.  

The DD-WRT control panel also provides an option to change the transmission 
power with which we can increase/decrease the signal strength. 
The normal signal strength for the router detected by our target device (a
MacBook Air laptop) was around -39 dBm.  The router and the target device were
located around 30 cm apart.
Merely by adjusting router settings, we were able to vary the signal strength
of the router, as sensed by the target device, between -25 dBm and -48 dBm.
By changing the distance between the target device and the spoofed router, we
were able to further reduce the signal strength down to -87 dBm.  This suggests
that the adversary has a high degree of control in manipulating sensed signal
strength.
Based on this spoofing and Received Signal Strength Indicator (RSSI) manipulation capability, the Wi-Fi context
matching attack becomes rather straightforward. The attacker can even have
advantage in environments where number of Wi-Fi APs is low. For example, we
observed that there are less than five APs in outdoors such as parking lot. In
such cases, the attacker would only need to spoof \prover 's side.

\subsection{Manipulating Physical Environment Sensor Modalities}
\label{sec:physical}
As discussed in \cite{ShresthaFC2014}, it may seem hard to manipulate physical
modalities, Temperature {\temperature}, Humidity {\humidity}, Gas {\gas} and Altitude  {\altitude}.
For example, it appears that an adversary has to change the temperature or
humidity of the entire environment surrounding the victim device which may be
quite challenging or detected easily.  However, in this section, we show
that, by using off-the-shelf devices, manipulating physical context is not only
feasible but also realistic and effective by tampering with the ``local''
environment close to one of the devices (e.g., an unattended \verifier). Our
attacks do not require the compromise of the devices (\verifier or \prover),
but rather only manipulation of environment close to their sensors. In order to
monitor the current ambient readings as they are being changed, the attacker
has to use his sensors. These ambient readings serve as a feedback for the
attacker while he attempts to change the current \verifier's ambience.  The
feedback sensor needs to be placed very close to the victim sensor so that the
two provide similar readings.


Our experiments demonstrate how different sensor modalities can be
\textit{manipulated, controlled} and \textit{stabilized} to enable successful
relay attacks. Arbitrarily changing a sensor's readings, at the verifier's
side, based on a physical activity may be straightforward but consistently
maintaining and controlling these readings to match those at the prover's side,
is non-trivial. For example, it may be obvious that temperature can be
increased using a hair dryer (a simple tool used in our temperature
manipulation experiments), but how to maintain it at a desired level for a
reasonable period of time (during which the attack can be launched) is not
obvious. While we present several direct/explicit ways to manipulate many
modalities, we also demonstrate some indirect/implicit techniques. For example,
we show how altitude can be manipulated by changing pressure (i.e., without
relocating the device to a different altitude).
%
%
%
%
When performing the attacks, we need to consider that the attacker will not have 
access to the direct readings from the actual (\verifier) device
and hence has to use his own sensors to monitor the current ambient readings during the attack.
These ambient readings serve as a
feedback for the attacker while he attempts to change the current \verifier's
ambience.  The feedback sensor needs to be
placed close to the victim sensor so that both provide similar readings.

\subsubsection{Temperature Manipulation}
\label{sec:temperature}
We were able to successfully alter the temperature to a desired level using
various household items, such as a hair dryer, a coffee mug, and ice cubes.
All of our experiments were performed with Sensordrone devices serving as both
\verifier and the attacker's feedback sensor.

\begin{figure}[htpb]
\minipage[b]{\linewidth}
\centering
 \includegraphics[width=1\textwidth]{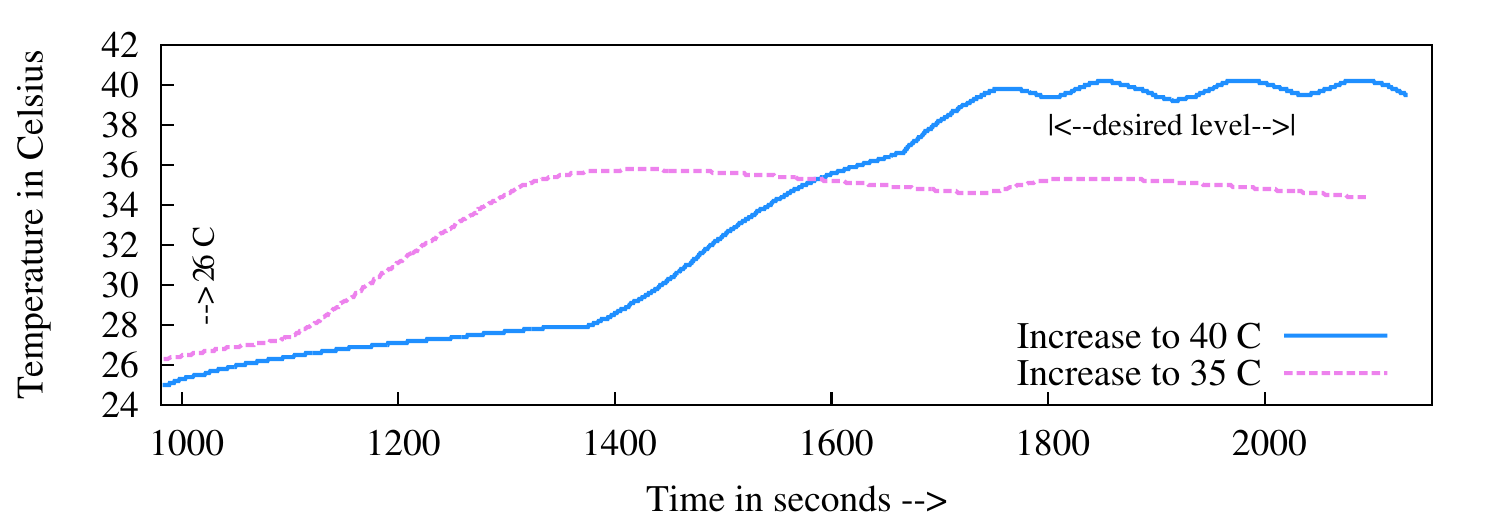}
  \caption{Increasing \temperature to desired level (35 \textdegree C and 40 \textdegree C)}
  \label{fig:TempIncrease35n40}
\endminipage
\end{figure}

\smallskip \noindent \textbf{Increasing the Temperature:} In
situations where \prover (e.g., a car key indoors) is at a higher temperature
than \verifier (e.g., a car parked outside in winter), the attacker must
increase the temperature.  We first used
a hair dryer to heat-up the area around the Sensordrone such that the
temperature is increased to a desired level.  To monitor how the temperature
increases as we bring the hair dryer closer to \verifier, we first placed the
hair dryer far enough and then brought the hair dryer closer to the sensors in
a way that we can handle the increase in temperature gradient.  
In our experiment, we first tried to increase the temperature to 40 \textdegree C 
and then to 35 \textdegree C. After a few attempts,
we could successfully increase the temperature to a desired level and stabilize
for almost 2 minutes (Fig. \ref{fig:TempIncrease35n40}). The lab
temperature when the experiments were performed was around 26 to 27 \textdegree
C. The hair dryer we used \cite{hairDryer} had a power of 1875 watt AC. A
video demonstration of our attack has been uploaded to YouTube~\cite{heatingVideo}.
\verOne{To perform this attack in a real world, the attacker can use a battery-operated 
device or a power outlet from his vehicle when \verifier is located in a parking lot.}

\begin{figure}[htpb]
\begin{minipage}[b]{\linewidth}
      \centering
      \includegraphics[width=1\textwidth]{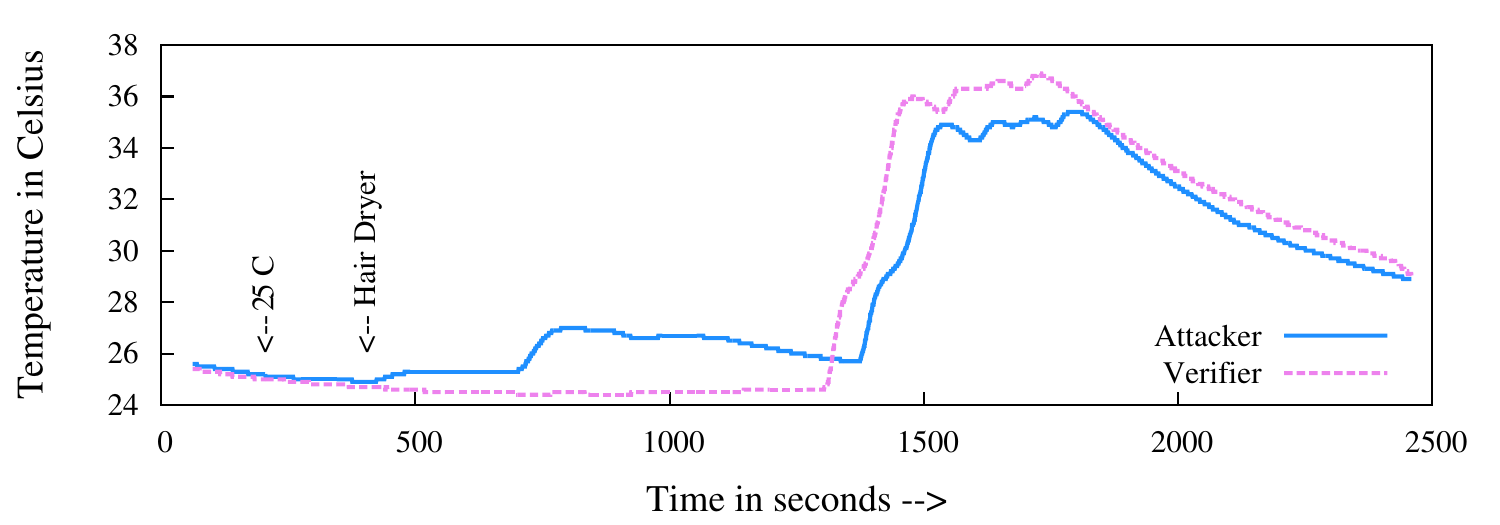}
      \caption{$VS$ and $FS$ on same location; the attacker trying to increase temperature to 35 \textdegree C.}
      \label{fig:HD_VASamePoint}
    \end{minipage}
    
\end{figure}

\begin{figure}[htpb]
    \begin{minipage}[b]{\linewidth}
      \centering
      \includegraphics[width=1\textwidth]{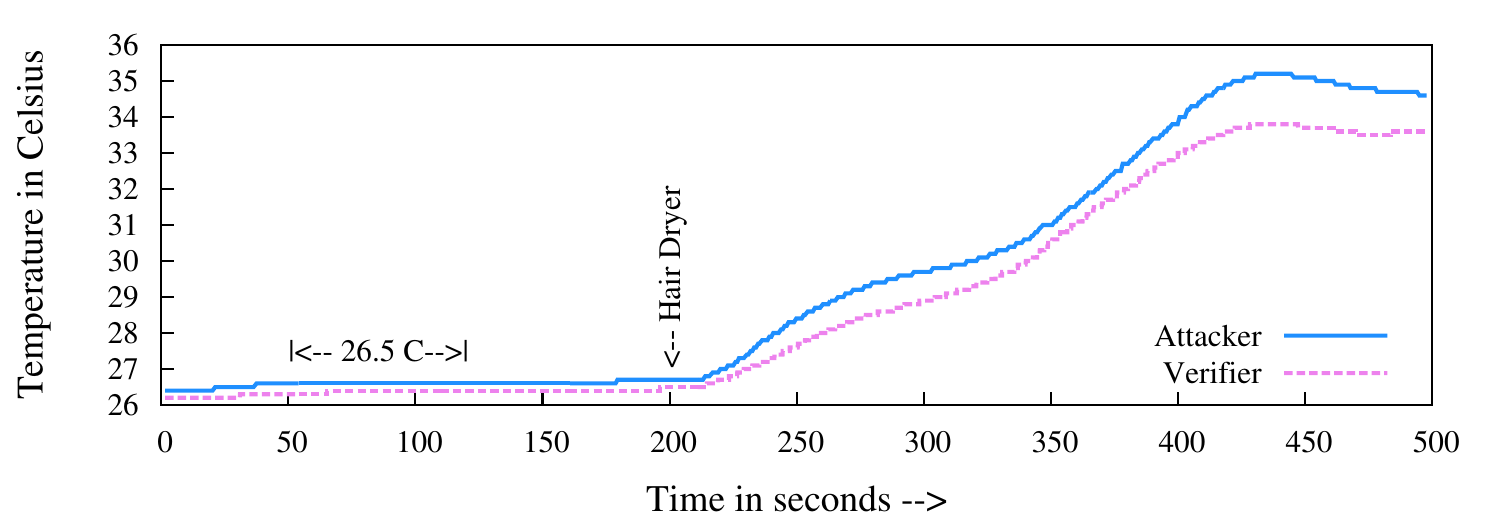}
      \caption{Heating an area; $VS$ and $FS$ within a range of 15 cm; the attacker trying to increase temperature to 35 \textdegree C.}
      \label{fig:HD_VA15cmRange}
    \end{minipage}%
\end{figure}


Our next set-up uses two sensors, \verifier sensor ($VS$) and feedback sensor ($FS$), to change the temperature.
Depending on whether or not the attacker knows where
the sensor is precisely located on \verifier device, he may place 
$FS$ either exactly on top of $VS$ or away from
it. 
We performed the hair dryer test such that: (1) $FS$ is placed at the same
place as $VS$; (2) $FS$ is placed such that $VS$ is closer to hair dryer than $FS$; and 
(3) $FS$ is placed such that $FS$ is closer to hair dryer than $VS$.

For the first case, 
we were able to match the temperature on both sensors to a large extent when
performing the heating activity (Fig. \ref{fig:HD_VASamePoint}).
However, if the attacker does not know the location of $VS$ then the sensor device closer
to the hair dryer ends up getting more heated.
These attacks are described in Appendix \ref{sec:a1} in detail.
Hence, the attacker should heat up the whole area as he may not be 
able to 
place his $FS$ exactly on top of $VS$. Subsequently, we tried to apply the heat not
just focusing on one particular area but rather heating the entire area within
a range of 15 cm.
%
%
Using this approach, we could effectively change
the temperature around $VS$ with feedback from  $FS$ as the two temperature
curves move side by side (Fig. \ref{fig:HD_VA15cmRange}). 
We were able to control the temperature to a desired level within a variance of
+/-0.3 \textdegree C for more than one minute in $FS$ device.
\verOne{It took us less than 4 minutes to reach the desired temperature level. We also observed that our ability to make the sensor reach the target temperature and sustain it for a significant duration (\texttildelow1 minute) increased dramatically with experience. This suggests that determined and professional attacker will fare significantly better than our results.}

\begin{figure}[htpb]
    \begin{minipage}[b]{\linewidth}
      \centering
      \includegraphics[width=1\textwidth]{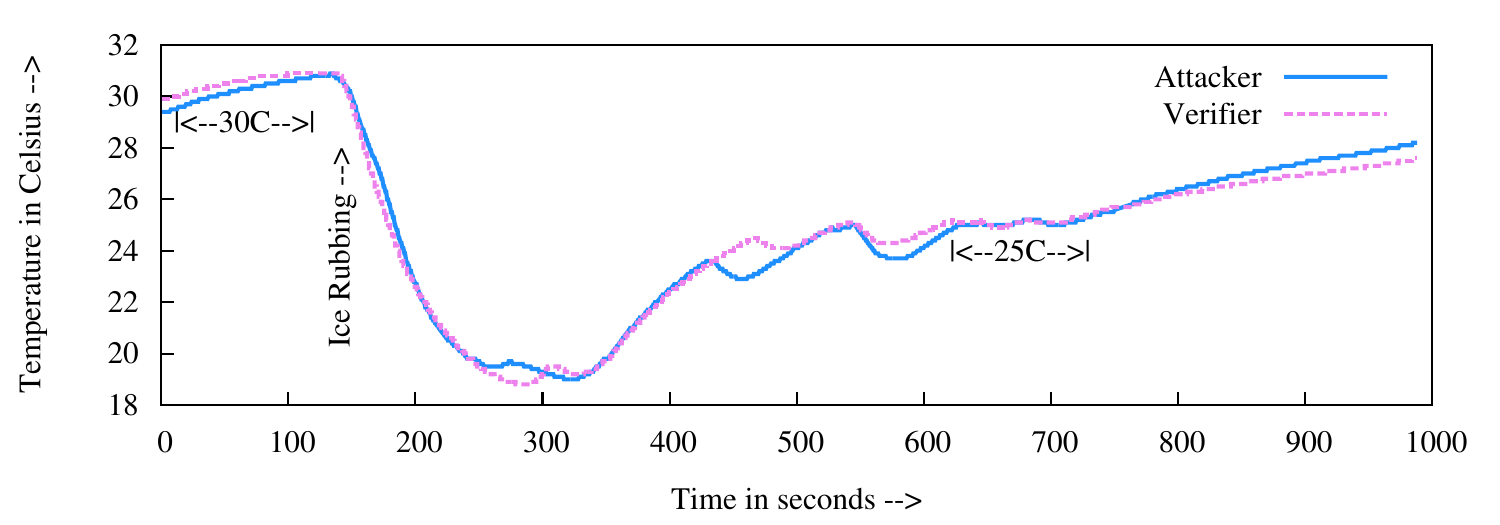}
      \caption{Decreasing temperature with an ice cube; the attacker trying to decrease to 25 \textdegree C.}
      \label{fig:TempDecreaseIce}
    \end{minipage}
\end{figure}

\smallskip \noindent \textbf{Decreasing the Temperature:} In some scenarios, it
might be necessary for the attacker to reduce the temperature recorded by
\verifier (e.g., when \prover is indoors and \verifier is outdoors during summer
conditions).
To decrease the temperature readings, we used an ice cube and rubbed it against
the sensor.  
The environment on the other hand increased the temperature. By using the ice cube, 
we first tried to drop the temperature below 20 \textdegree C
and then let the environment increase the temperature naturally. This natural
increase of the temperature was very slow, and when the temperature started
increasing beyond the desired temperature level, we gently rubbed the ice again
to stabilize the temperature. 
We conducted experiment in a parking deck where the ambient temperature was
around 30 \textdegree C. Our goal was to change the temperature
down to 25 \textdegree C.  We rubbed the ice cube on the sensors (both
\verifier and feedback sensors) until the temperature decreased to less than 20
\textdegree C.  Afterwards, the temperature started rising slowly naturally.
When it reached around 25.2 \textdegree C, the ice cube was rubbed gently again
on the sensors such that the temperature drops slightly. We were able to
decrease the temperature and stabilize it at 25 \textdegree C for more than a
minute after a few trials within a variance of +/-0.3 \textdegree C as shown in Fig.~\ref{fig:TempDecreaseIce}.

\subsubsection{Humidity Manipulation}
\label{sec:humidity}
To alter humidity, we used common household items 
such as hot coffee (for increasing humidity) and hair dryer (for decreasing humidity).

\begin{figure}[htpb]
    \begin{minipage}[b]{\linewidth}
        \centering
        \includegraphics[width=1\textwidth]{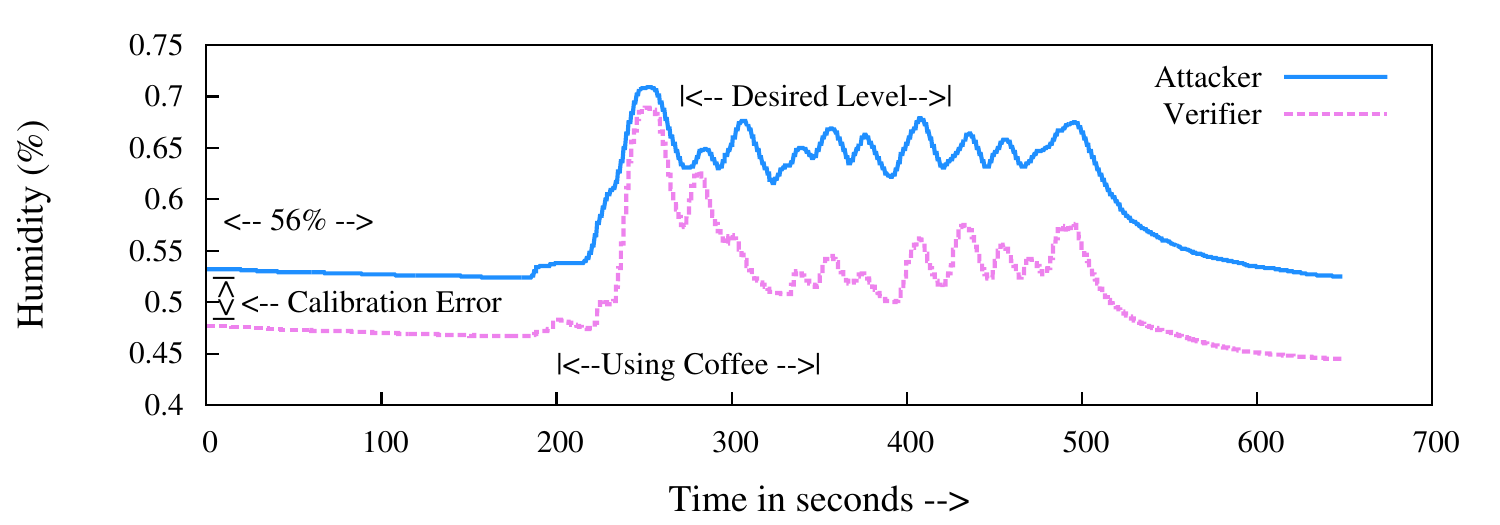}
        \caption{Increasing humidity with hot coffee; the attacker trying to increase to 65\%.}
        \label{fig:HumidityCoffee}
    \end{minipage}%
\end{figure}

\smallskip
\noindent \textbf{Increasing the Humidity:} 
Coffee fumes when brought close to $VS$ would increase the humidity level. 
An attacker has to move the hot coffee cup
nearer to, and farther away, from the sensors to control the humidity level. 
Using this strategy, we were able to increase the humidity by
10\%, i.e., from normal humidity of 55\% to 65\% (Fig.
\ref{fig:HumidityCoffee}).  
The attacker needs to use $FS$ to control the humidity. On our first
attempt, we were able to control the humidity with a variance of +/-3\% for
almost 30 seconds. In the second attempt, we could raise the humidity to the
desired level for more than one minute (106 seconds) with the same threshold.


\smallskip

\begin{figure}[htpb]
      \vspace{-5mm}
    \begin{minipage}[b]{\linewidth}
      \centering
      \includegraphics[width=1\textwidth]{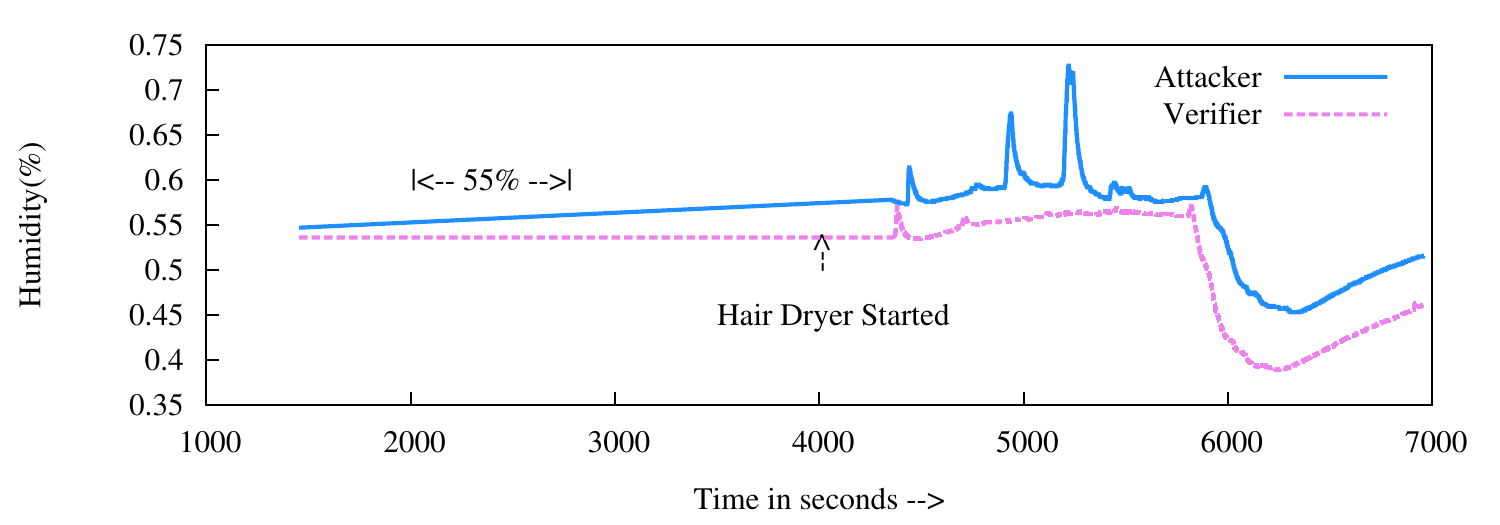}
      \caption{Decreasing humidity with hair dryer such that $VS$ and $FS$ are at same location; the attacker trying to decrease to 50\%.}
      \label{fig:HD_HumidSameLoc}
    \end{minipage}
\end{figure}

\begin{figure}[htpb]
      \vspace{-5mm}
    \begin{minipage}[b]{\linewidth}
      \centering
      \includegraphics[width=\textwidth]{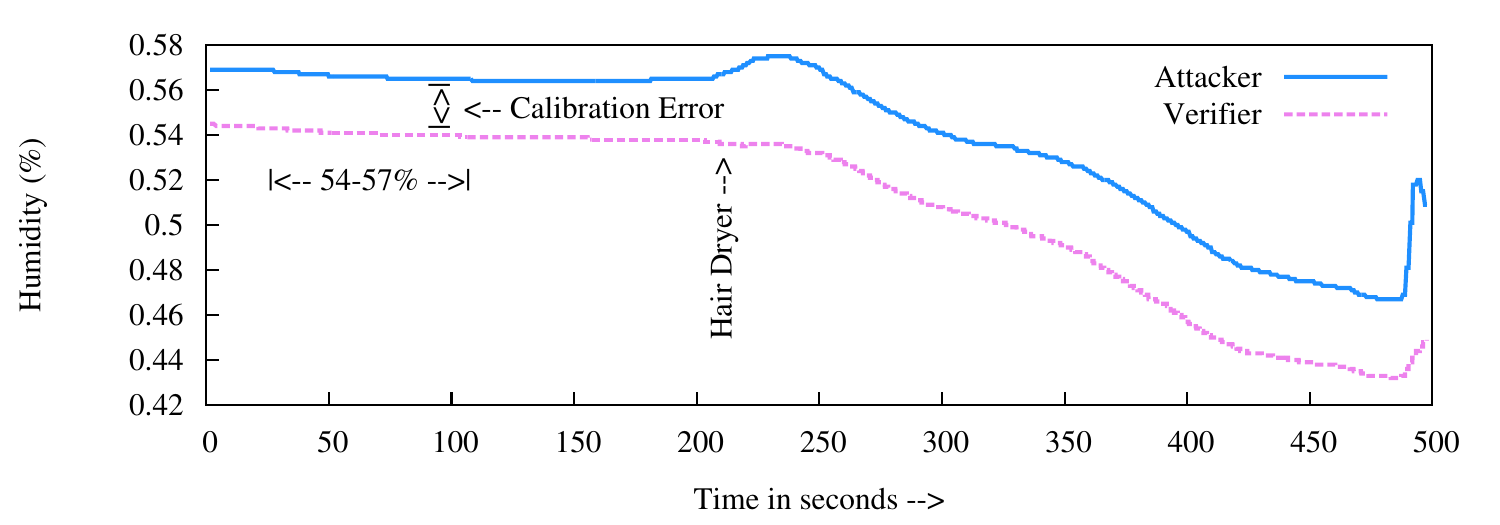}
      \caption{Decreasing humidity with hair dryer such that $VS$ and $FS$ are within a range of 15 cm; the attacker trying to decrease to 50\%.}
      \label{fig:HD_Humid15cm}
    \end{minipage}%
\end{figure}

\noindent \textbf{Decreasing the Humidity:} 
A hair dryer can be used to dry-up
the air around the sensor to reduce the humidity. 
The setup of this experiment is similar to the hair dryer temperature increase
experiment.  We tried to decrease the humidity of $VS$ by monitoring the
humidity change on $FS$. When two devices are placed exactly at the same
location, the humidity decreases and matches consistently between the two
devices (Fig. \ref{fig:HD_HumidSameLoc}). Even when the two devices are placed 15 cm apart,
the drop in the humidity readings coincides (Fig. \ref{fig:HD_Humid15cm}). 

\subsubsection{Gas Manipulation}
\label{sec:gas}

Following prior work \cite{ShresthaFC2014}, we study Carbon Monoxide (CO) level as a modality for
co-presence detection.
While manipulating this modality, an attacker may
not be detected even when he alters the gas content near either \verifier or
\prover (unlike the rudimentary model of Section
\ref{subsec:threatmodel}),
unless there is a significantly large change, or gas monitors are installed.
This provides flexibility to the attacker to increase/decrease the CO level at
both sides such that both readings match.


\smallskip

\noindent \textbf{Increasing the Gas (CO):} We performed several activities
such as using a smoking cigarette to exhale a high amount of CO gas to the
sensor, and using a car exhaust to increase the CO level.  We also found out
that room heaters emit gases which increase CO readings when we placed the
sensor device on top the gas vent while the heater was turned on. The aerosol
spray 
also increased the CO
level when it was sprayed around on top of the sensor. The effect of different
propane gas heaters as well as aerosols air fresheners on gas content has been
mentioned in \cite{NYCFireSafetyCOMonitor}. 
All these activities, though, increased the CO level abruptly, 
it takes a long time for sensor reading to descend back to normal, 
which provides the attacker with a sufficiently long attack window as shown in Fig. 
\ref{fig:gas_freshener}. 
The effects of cigarette and car exhaust on CO level are described in Appendix \ref{sec:a2} in detail.
We observed these activities for more than five times, and
noticed that it took more than thirty seconds to decrease by 1 ppm when gas
level decreased below 10 ppm which is already above average of normal gas
level.

\begin{figure}[htpb]
    \begin{minipage}[b]{\linewidth}
      \centering
      \includegraphics[width=\textwidth]{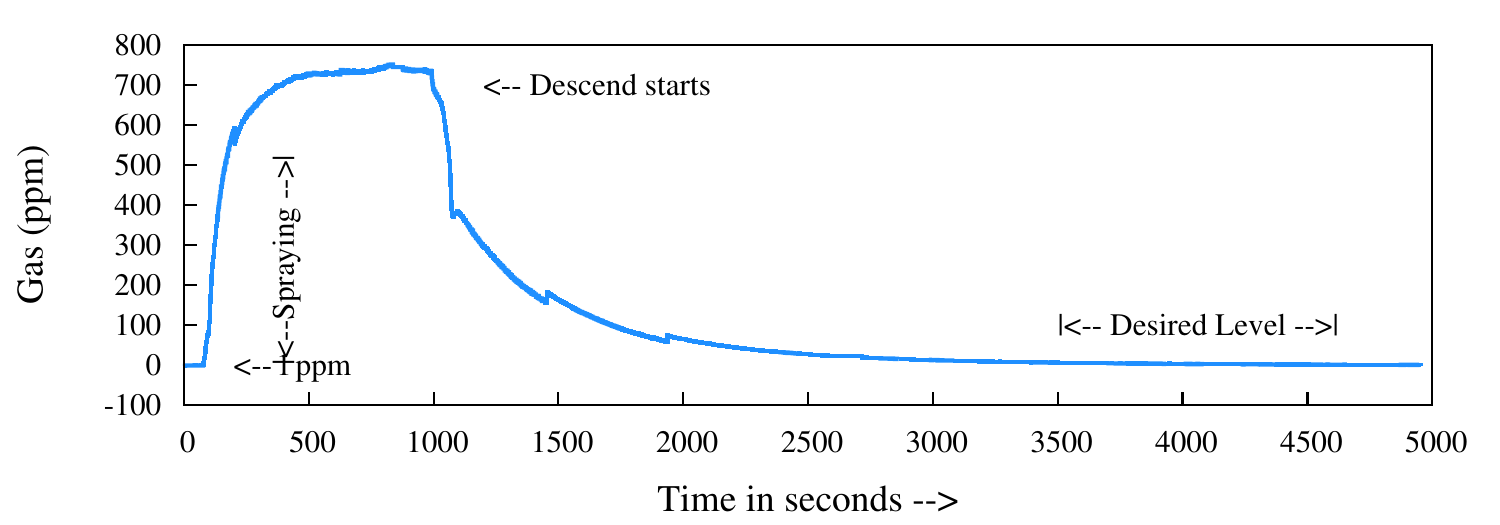}
      \caption{Effect of aerosol spray in CO level; increasing the CO gas 
      level to arbitrary value and wait to decrease to desired level.}
      \label{fig:gas_freshener}
    \end{minipage}
\end{figure}


\smallskip

\noindent \textbf{Decreasing the Gas (CO):} 
To reduce the gas level, an attacker needs to ``purify'' the air from the CO
content around the sensors. We implemented this strategy using a kitchen
exhaust fan which is used to remove pollutants. We found that when sensor was
placed near the exhaust fan, it decreased the CO gas content. 

The gas reading heavily depends upon the location of \prover and \verifier.  In
a heavy traffic or polluted area, this may be higher than 10 ppm while in a
normal workplace, it may be around 0 ppm to 5 ppm. If \prover is located in
low CO area while \verifier is located in high CO area, the attacker may use the
kitchen exhaust fan activity to decrease the CO level in \verifier's location.
However, if the attacker cannot reduce the CO level by significant amount, he can
always collude with the attacker at \prover's side to increase the CO level
using an aerosol spray. This can increase the CO level by significant amount and then
it only takes a while to fall back to the normal gas level. This effect can be
confirmed from  Fig. \ref{fig:gas_freshener}.

\subsubsection{Altitude Manipulation}
\label{sec:altitude}
The altitude of a location is inversely correlated to the pressure at that location. The 
Sensordrone device detects the pressure, and uses it to calculate the altitude based on a 
standard conversion method. 

Manipulating sensors so as to increase or decrease altitude directly seems very
difficult.  In order to manipulate the altitude readings, one may physically
carry the verifier device to a higher or lower altitude as needed. If the
verifier device is portable (such as a stolen laptop), doing so is easy. 
However, there are many scenarios where directly changing the altitude is not feasible (e.g., when \verifier is a car and \prover is a
car key carried in victim's pocket).  We show that it is still possible to
manipulate altitude readings \textit{indirectly} by manipulating the pressure
readings.

\begin{figure}[htpb]
    \begin{minipage}[b]{\linewidth}
      \centering
      \includegraphics[width=1\textwidth]{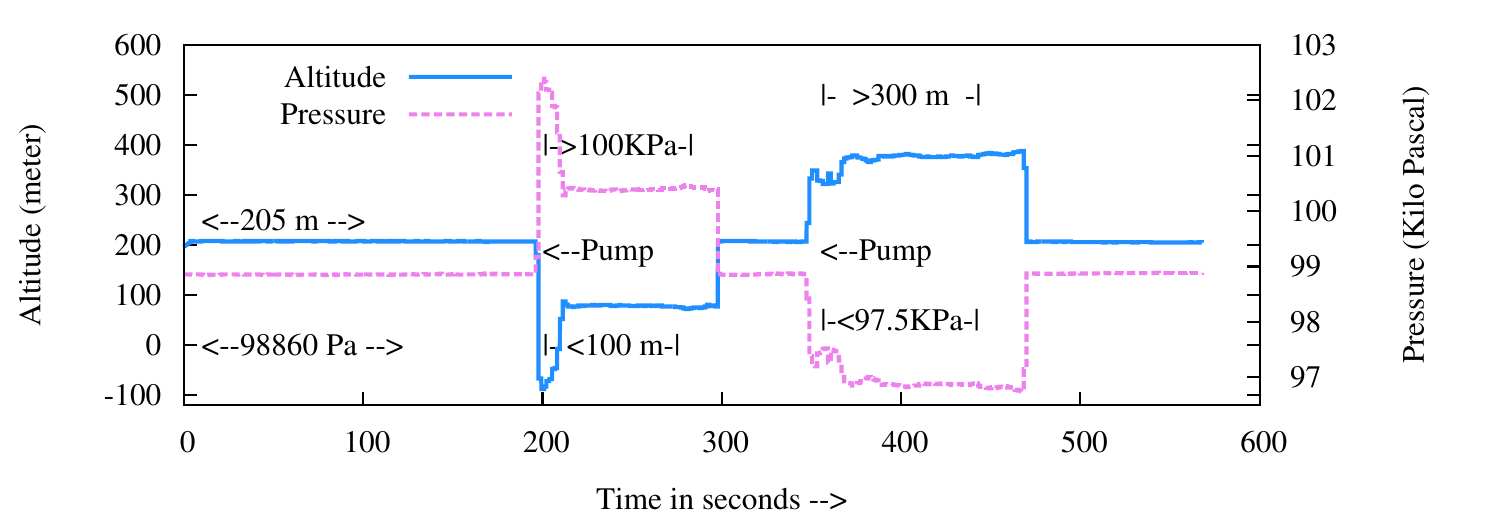}
      \caption{Using an air pump to change pressure to the sensors wrapped inside a Ziploc bag by pumping air in and out.}
      \label{fig:ziploc}
    \end{minipage}%
\end{figure}

\smallskip \noindent \textbf{Increasing the Altitude:} 
To increase the altitude indirectly, an attacker must decrease the pressure
near the sensors.  To achieve this functionality, we created a low-cost air compressor.  We
placed the sensor inside a Ziploc bag and then used an electric air pump~\cite{airpump} 
to suck-up the air from the bag. 
When \verifier is large in size or shape (such as a car), an attacker just needs to 
create an enclosure around its sensor, while if it is a portable/small device 
(e.g., a laptop), the device itself can be placed inside a bag.
When the air pump sucks up the air around the
sensors enclosed inside the Ziploc bag, the weight of air exerted on the sensor
is reduced. This reduces the pressure around the sensor and hence increases the
altitude level. In our experiment, we effectively altered the altitude by more than 60
meters (Fig. \ref{fig:ziploc}). By using an air pump with a higher power, 
the attacker can further increase the altitude level.
A vacuum cleaner may also be used in place of an air pump (as described in Appendix~\ref{sec:a3}). 

\smallskip \noindent \textbf{Decreasing the Altitude:} To decrease the altitude
(i.e., increase the pressure), we placed the sensor inside a polythene bag and
applied high pressure by squeezing the bag, blowing air into the bag, and
finally using the air pump device to blow the air inside. 
First, we wrapped the sensor inside a polythene bag to see if there is any change in
altitude when we blow air into the bag by mouth, or squeeze the air tight polythene bag.
This increased the pressure by very high amount and decreased the altitude
correspondingly. However, it was not doable in a controlled way, i.e., sometimes the altitude 
decreased by 5 meters while on other occasions, it decreased by 50 meters. Ideally, an 
attacker would want to have a relatively long time window where the desired altitude remains 
constant for him to perform the relay attack. To address this issue, we used the air pump 
mentioned above. Filling up the air into the bag increased the pressure and decreased the 
altitude such that it remained constant for almost 14 seconds. A video demo of this experiment 
has been uploaded to YouTube \cite{ziplocVideo}.


\subsection{Manipulating Multiple Sensor Modalities Simultaneously}
\label{sec:multiple}

As demonstrated by prior work
\cite{DBLP:conf/esorics/HaleviMSX12,TruongPerCom14,ShresthaFC2014}, a
contextual co-presence detection system can use 
combinations of several sensor modalities. 
In such cases, the attacker needs to manipulate multiple modalities at the same
time (\multimodal attacker). 
However, performing one
activity may be altering not only the target modality but also 
one or more other modalities 
that a system might be using for context detection, such as (\temperature
and \humidity) or (\altitude and \audio) even though they are not directly
correlated. 

For example, hair dryer increases temperature but also dries-up the air (i.e.,
potentially reduces the humidity) around the sensor where it is applied. It
also changes the ambient noise. An attacker needs to manipulate in such a way
that if the multiple modalities are involved in the system he should change the
target modality without altering other modalities by effective amount.
We also found that hair dryer activity results in a huge momentary change in
gas level. However, the reading comes back to normal when hair dryer
is applied for a long period of time. Altitude and pressure did not change with
the hair dryer activity. Hair dryer activity also does not impact on RF
signals. Hence, hair dryer activity can be used to manipulate the system which
uses either temperature or humidity along with gas, altitude and RF signals.

Using aerosol spray to increase the gas content does not have effective change on
any other modalities besides humidity. Similarly, updating RF signals does not
seem to have any effect on physical modalities. Therefore, an attacker can
simultaneously manipulate radio, temperature and gas while he hopes that audio,
altitude and humidity either match the minimum criteria from both sides or is
not used by the system.

Using an ice cube to decrease the temperature does not affect other modalities
effectively. 
However, if the ice melts then
it may affect the humidity of the space near the sensors. In our experiment, we
saw that humidity fluctuates when we tried to decrease the temperature using an
ice cube. Hence, using an ice cube to decrease temperature activity can be used
with all other modalities except altitude and humidity.

Hot coffee cup changes the humidity along with the temperature,
while other modalities remain unchanged. In this case, an attacker can
manipulate humidity along with radio, audio and gas while he cannot control
temperature and humidity together.

When an attacker has to use an air pump or vacuum cleaner to increase or
decrease the altitude, it affects ambient noise. Also, an air pump was used in
conjunction with a Ziploc bag where the sensors were wrapped to create an
enclosed space.  When an attacker performs such activity with an enclosed
space, it will be very difficult for him to change gas, temperature or
humidity. We thus may only claim that the attacker can manipulate altitude
along with radio modalities.

To summarize, our attacks support the following combinations of \multimodal manipulations:
(1) \altitude, \bluetooth, \wifi;
(2) \audio, \bluetooth, \gas, (increase for \humidity), \wifi;
(3) \audio, \bluetooth, \gas, (decrease for \temperature), \wifi;
(4) \audio, \bluetooth, \gas, \wifi;
(5) \bluetooth, \gas, \humidity, \wifi;
(6) \bluetooth, \gas, \temperature, \wifi. 
However, a more sophisticated attacker (than the one we considered) may use 
different techniques to possibly attack other combinations too.

\section{Security of Co-Presence Detection Systems Under Context Manipulation}
\label{sec:analysis}

In light of the attacks presented in
Section~\ref{sec:manipulation_attack}, we first extend the rudimentary
contextual attacker model from \cite{TruongPerCom14} as follows:

\begin{itemize} \itemsep0em

\item We allow \multimodal attackers who can simultaneously control
multiple sensor modalities, in addition to the \singlemodal attacker of
\cite{TruongPerCom14}. 

\item 
We assume that a contextual attacker can
manipulate radio contexts in \emph{both directions}. The same 
assumption applies to Gas sensors in light of our aerosol spray attack.
\end{itemize}


\subsection{Analysis Methodology} \label{sec:analysis_method}

\verOne{ 

\subsubsection{Datasets}

To fairly evaluate the resilience of co-presence detection systems in the
presence of our contextual attacker, we used the same datasets and the same set
of features originally used to evaluate the systems in question. The previous
audio-radio system \cite{TruongPerCom14} used a dataset 
to evaluate resistance against \singlemodal attackers. The previous physical system
\cite{ShresthaFC2014} used a dataset 
to model a \zeromodal attacker. We use these datasets to evaluate the
resistance of the respective systems against \multimodal attackers.  In
addition, 
we conducted new audio relaying experiments to collect data and evaluate audio-based co-presence detection performance.
Furthermore, we collected a new dataset corresponding to the
audio-radio-physical system (which was not considered in prior works). 

\textbf{Data-PerCom}. The dataset from Truong et al. \cite{TruongPerCom14} contains 2303 samples, of which 1140 samples (49.5\%) are from co-present devices and 1163 (50.5\%) from non co-present devices. Each sample contains data from sensor modalities available at the time on the respective devices (2117 with audio, 1600 with Bluetooth, 782 with GPS and 2269 with Wi-Fi).  Recording time varies from sensor to sensor: 2 minutes for GPS scanning, 10 scans for Wi-Fi (about 30 seconds), 10 seconds for ambient audio, and 10 scans for Bluetooth (up to 12 seconds for each scan).

\textbf{Data-FC}. The data from Shrestha et al. \cite{ShresthaFC2014} contains sensor data 
for ambient temperature, precision gas, humidity, and 
altitude. Data was recorded and labeled according to the location and time
of the place. The data was also marked how the device was held, i.e.,
either in hand or in pocket. The experiment was conducted in a variety of
places, not just confined to labs and typical university offices. The 
locations included: parking lots, office premises, restaurants,
chemistry labs, libraries as well as halls with live performance and 
driving on interstate highways. We collected a total of 207 samples at 21
different locations. The different samples collected from the same place are
``paired'' to generate co-presence data instances whereas those from different
places are paired to generate non co-presence data instances. We ended up with
21320 instances of which 20134 instances belonging to non co-presence class and
1186 instances belonging to co-presence class.

\textbf{Data-TMC-audio}. To assess how an attacker can manipulate ambient audio via the
streaming attack (Section \ref{sec:audiomanipulate}), we conducted a
set of experiments to collect about 100 audio samples for the non
co-presence case. The audio streaming was done over two different channels:
Wi-Fi and cellular data. \prover was a Galaxy Nexus device while
\verifier was a Galaxy S3 device. Unidirectional streaming of the audio from
\prover's side to \verifier's side was done between a pair of devices (from
a Galaxy S4 to an iPhone 5 in the case of the cellular data channel,
and from a MacBook Air to a ThinkPad Carbon X1 in the Wi-Fi
channel). The attacker devices used a Skype connection as the audio
relay channel.

\textbf{Data-TMC}. We extended the data collector used in \cite{TruongPerCom14} to record physical
sensor data using an attached Sensordrone device (as used in
\cite{ShresthaFC2014}). Different device models were used to record sensor
data. Each device, in a pair of devices, was connected to its own Sensordrone
device. Two users were involved in the data collection. Data was collected at
different locations in two countries for ten days. The resulting dataset has
203 non co-presence samples and 335 co-presence samples.

%


\subsubsection{Features for co-presence detection}
We summarize features used for co-presence detection in prior works.

\textbf{Audio features}. Halevi et al. \cite{DBLP:conf/esorics/HaleviMSX12} proposed the use of (only) audio
for co-presence detection. Audio features include max cross correlation and time frequency distance.

\begin{itemize}
\item Max cross correlation: \\ $ M_{corr}(a,b) = Max(cross\ correlation(X_a,X_b))$
\item Time frequency distance: \\ $D(a,b) = \sqrt{(D_{c,time}(a,b))^2 + (D_{d,freq}(a,b))^2}$
where, $D_{c,time}(a,b) = 1 - M_{corr}$, $D_{d,freq}(a,b)$ =
$||FFT(X_a) - FFT(X_b)||$ is the Euclidean norm of the distance.
\end{itemize}
Here $X_a$ and $X_b$ denote the raw (16-bit PCM) audio signals recorded by $A$ and $B$ and FFT($X_a$),
FFT($X_a$) denotes the Fast Fourier Transforms of the corresponding signals.

\textbf{Radio (Bluetooth, Wi-Fi, GPS) features}. Truong et al. \cite{TruongPerCom14} used a set of features with small variance for radio sensor modalities. 
They let a sample from an RF sensor modality be of
the form ($m$, $s$) where $m$ is an identifier of a sensed device and
$s$ is the associated signal strength.  Also, they let $S_a$ and $S_b$ denote the
set of records sensed by a pair of bound devices $A$ and $B$, and let
$n_a$ and $n_b$ denote the number of different beacons (i.e., Wi-Fi
access points, satellites or Bluetooth devices) observed by devices
$a$ and $b$. We define the following sets:

\noindent $S_{a} = \{(m^{(a)}_i,s^{(a)}_i)\ |\ i \in
\mathbb{Z}_{n_a-1} \}$. \\
\noindent $S_{b} = \{(m^{(b)}_i,s^{(b)}_i)\ |\ i \in
\mathbb{Z}_{n_b-1} \}$. \\
\noindent $S_a^{(m)} = \{m~ \forall (m,s) \in S_a\}$, $S_b^{(m)} = \{m~ \forall (m,s) \in S_b\}$. \\
\noindent $S_{\cap}$ = $\{(m,s^{(a)},s^{(b)})$ $\forall m |(m,s^{(a)}) 
\in S_{a}, (m,s^{(b)}) \in S_{b} \}$. \\
\noindent $S_{\cup}$ = $S_{\cap}$ $\cup$ $\{(m,s^{(a)},\theta)
~\forall m | (m,s^{(a)}) \in S_a, m \not \in S_b^{(m)} \}$ $\ \\\hspace*{1.45cm}\cup $
$\{(m,\theta, s^{(b)}) ~\forall m | (m,s^{(b)}) \in S_b, m \not \in
S_a^{(m)} \}$, \\\hspace*{1.3cm} $\theta$ is modality-specific (see below). \\
\noindent $S_{\cap}^{(m)} = \{ m ~\forall m | (m,s^{(a)},s^{(b)}) \in
S_{\cap}\}$. \\
\noindent $S_{\cup}^{(m)} = \{ m ~\forall m | (m,s^{(a)},s^{(b)}) \in
S_{\cup}\}$. \\
\noindent $L_{a}^{(s)} = \{ s^{a} | (m,s^{(a)},s^{(b)}) \in
S_{\cap}\}$. \\
\noindent
$L_{b}^{(s)} = \{ s^{b} | (m,s^{(a)},s^{(b)}) \in S_{\cap}\}$.

$S_{\cap}$ consists of devices seen by both $A$ and $B$;
$S_{\cup}$ represents all devices seen by $A$ or $B$ with
$\theta$ filled in as the ``signal strength'' for devices that are
\textit{not} seen by either device.  
\begin{enumerate}

\item \label{eq:jaccard}Jaccard distance: $1-\frac{|S^{(m)}_{\cap}|}{|S^{(m)}_{\cup}|}$

\item \label{eq:hamming}Mean of Hamming distance: $\frac{\sum\nolimits_{k=1}^{|S_{\cup}|} |s_k^{(a)} -
      s_k^{(b)}|}{|S_{\cup}|} $

\item \label{eq:euclid}Euclidean distance: $\sqrt{\sum\nolimits_{k=1}^{|S_{\cup}|} (s_k^{(a)} -
      s_k^{(b)})^{2}} $

\item \label{eq:expdif}Mean exponential of difference: $\frac{\sum\nolimits_{k=1}^{|S_{\cup}|} e^{|s_k^{(a)}
        - s_k^{(b)}|}}{|S_{\cup}|}$

\item \label{eq:sumsqranks}Sum of squared of ranks: $\sum\nolimits_{k=1}^{|S_{\cap}|} (r_k^{(a)}-r_k^{(b)})^2$


where,
$r_k^{(a)}$ (respectively $r_k^{(b)}$) is the rank of $s_k^{(a)}$
($s_k^{(b)}$) in the set $L_a$ ($L_b$) sorted in
ascending order.


\item \label{eq:subsetcnt} Subset count: $\sum\nolimits_{i=1}^{T} f_i$.
 Here $T$ is the scanning time (seconds)

 $f_i=1$ if $S_{a_i}^{(m)} \ne \emptyset$, $ S_{b_i}^{(m)} \ne
 \emptyset$, \\\hspace*{1.4cm}($S^{(m)}_{a_i} \subseteq S_{b_i}^{(m)} ~or~ S^{(m)}_{a_i} \supseteq S_{b_i}^{(m)}$)

 $f_i=0$ otherwise. $S_{a_i}$, $S_{b_i}$ are the set of records by $A$
 and $B$ respectively at the $i^{th}$ second

\end{enumerate}



Features~\ref{eq:jaccard}-\ref{eq:sumsqranks} are used for Wi-Fi ($\theta$ is -100.). Features~\ref{eq:jaccard}-\ref{eq:euclid} are used with BDADDR as 
identifier ($m$) and average RSSI as signal strength ($s$) for Bluetooth($\theta$ is -100). 

\textbf{Physical sensor features}. Shrestha et al. \cite{ShresthaFC2014} used Hamming distance as a feature on physical sensors for co-presence. 
Let $L_i$ and $L_j$ be a sensor reading captured by two devices at
locations $i$ and $j$. The Hamming distance is calculated as follows:

\begin{center}
$D(i,j)=|L_i-L_j|$
\end{center}

Given n different sensor modalities and the input data for the k$^{th}$ modality
$L_i^{(k)}$ and $L_j^{(k)}$ from
two samples, we have $D^{(k)}(i,j)=|L_i^{(k)}-L_j^{(k)}|$. With the 
data corresponding to $n$ modalities, we obtain a feature vector of
$n$ elements of $D^{(k)}(i, j)\ |\ 1\ \le k \le n$.

\subsubsection{Classification techniques}

In evaluating prior systems, we used the same classification techniques as
in the original evaluations (Decision Tree and Random Forest), implemented in
Scikit-learn~\cite{scikit-learn}. The results are reported after running
ten-fold cross validation. 
We use \textit{False Positive Rate (FPR) as a metric to 
represent the attacker's success probability}.  FPR corresponds to ``non
co-presence'' samples which are mislabeled as ``co-presence'', reflecting the
security of the system (higher the FPR, lower the security).  We use False
Negative Rate (FNR) as a metric to represent the usability of the system. FNR
represents ``co-presence'' samples that are mislabeled as ``non co-presence''
(lower the FNR, better the usability).  F1 score is reported only for the
overall performance of the classification model under \zeromodal attack.

Whenever multiple sensor modalities are used, we fuse the data from these
modalities before feeding it to the classifier. We considered the following
fusion approaches.


\textbf{Features-fusion}. The features of all sensor modalities are together fed to the classifier (see Fig. \ref{fig:feature-fusion}). The
decision of co-presence or non co-presence is made one-time only based on the output of the prediction model. Prior work \cite{TruongPerCom14,ShresthaFC2014} implemented this fusion technique.

\begin{figure}[htpb] 
	\centering
	\includegraphics[scale=0.2]{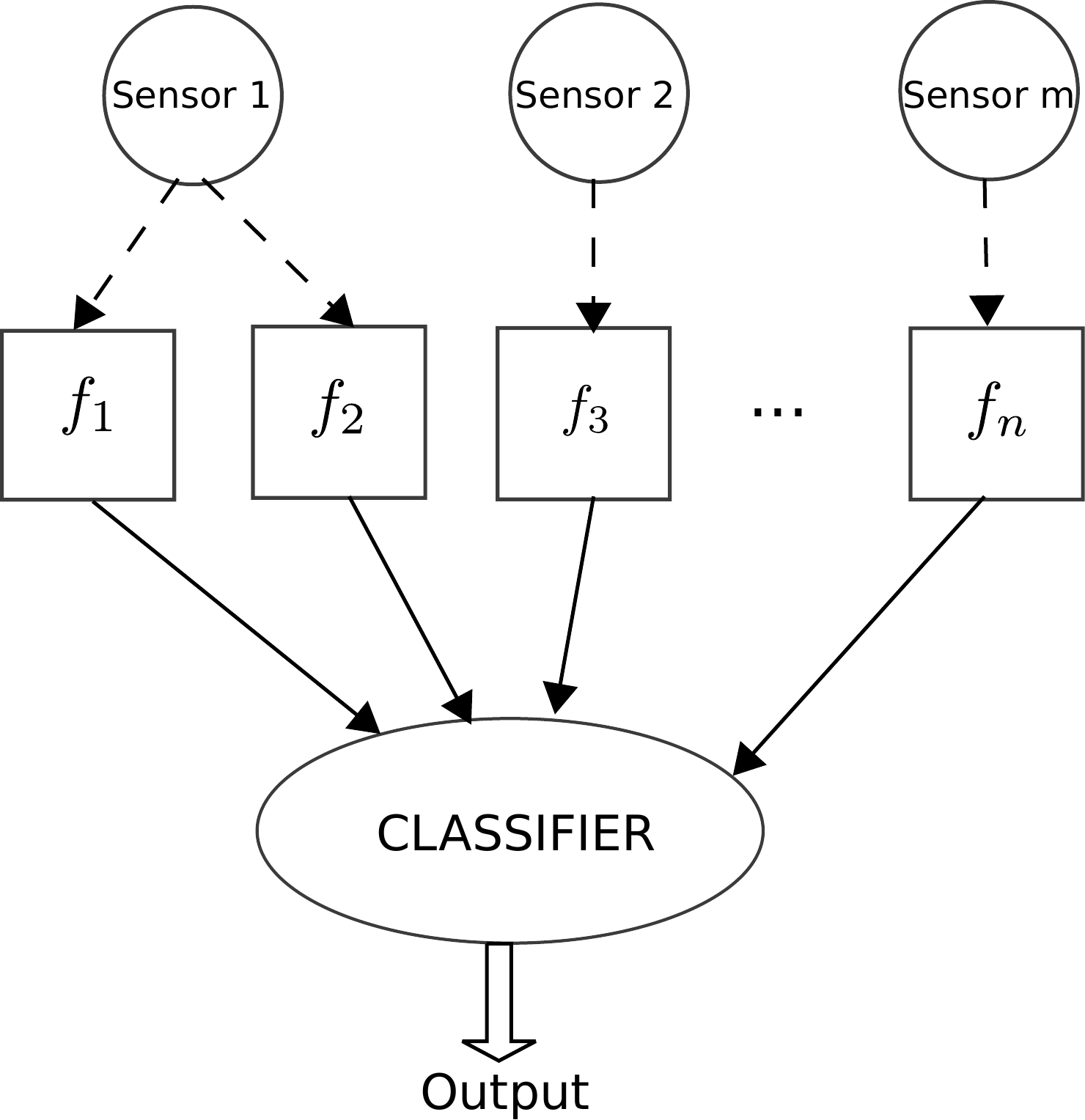}
	\caption{Features-fusion technique.}
	\label{fig:feature-fusion}
\end{figure}

\textbf{Decisions-fusion}. Each of the $n$ sensors (with all its features) is used separately by the
classifier. As result there are $n$ decisions made.  All decisions are then
combined to produce a final decision. This is an approach that has not been
used for co-presence detection in previous works. Decisions-fusion can
aggregate decisions from single sensor modalities or from subsets of sensor
modalities, for example, three subsets can be built on top of seven sensors:
acoustic = \{\audio\}, radio = \{\bluetooth, \wifi\}, physical = \{\altitude,
\gas, \humidity, \temperature\}. In the latter fusion approach, classifiers of
subsets are built using features-fusion. Fig. \ref{fig:decision-fusion} illustrates decision-fusion technique.

\begin{figure}[htpb] 
	\centering
	\includegraphics[scale=0.2]{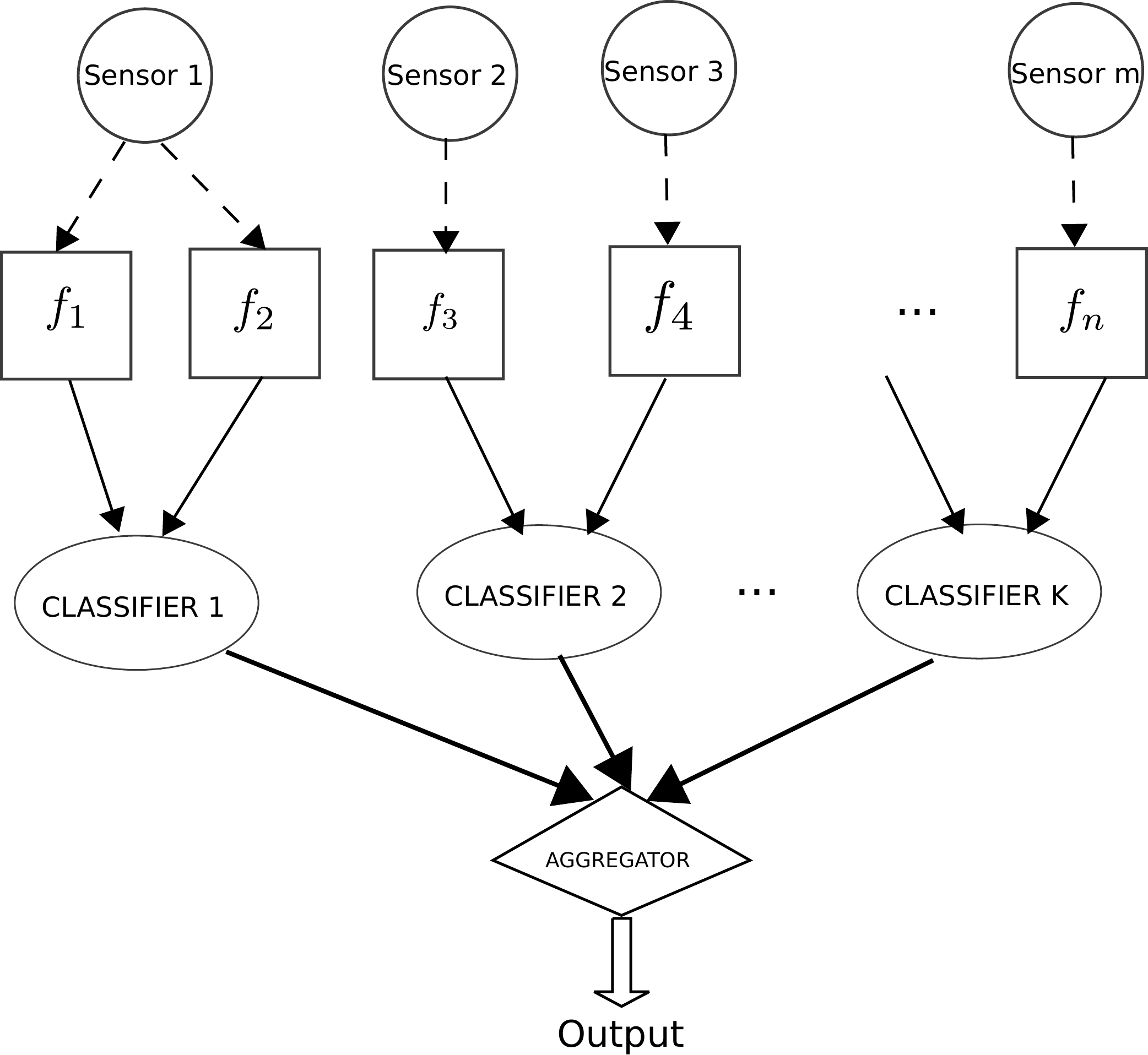}
	\caption{Decisions-fusion technique.}
	\label{fig:decision-fusion}
\end{figure} 

} 

\subsection{Analysis Results}

\subsubsection{Audio-Only System}
\label{sec:audio-only}
Halevi et al. \cite{DBLP:conf/esorics/HaleviMSX12} proposed the use of (only) audio
for co-presence detection. Their work showed that audio is a good ambient
context resulting in 100\% accuracy and 0\% False Positive Rate (FPR).

The audio features used in \cite{DBLP:conf/esorics/HaleviMSX12} are based on
audio frequency. Therefore, to evaluate the impact of frequency on the attack
feasibility, we tested three different ranges of ambient audio frequencies
collected by controlled experiments where we set up the ambient noise
surrounding recording devices falling into different categories. \textit{Low
ambient audio} (frequency less than 100 Hz); \textit{Medium ambient audio}
(frequency in the human audible range, at around 500 Hz); \textit{High ambient
audio} (frequency 5000 Hz or more).

\begin{table}[htbp]
    \centering
	\scriptsize
    \caption{Relay attack success rate (FPR) for audio streaming via Wi-Fi and
    Cellular networks} 
    \label{tbl:audio-streaming}

    \begin{tabular}{@{}lcrcr@{}}
        \toprule
        {\bf Acoustic relaying environments} &\phantom{} & {\bf Wi-Fi} &\phantom{} & {\bf Cellular} \\
	(\prover  freq $\rightarrow$ \verifier  freq) &\phantom{} & {} &\phantom{} & {} \\
        \midrule
        {\em High $\rightarrow$ Medium} && 100\% && 40\% \\
        {\em High $\rightarrow$ Low} && 100\% && 20\% \\
        {\em Medium $\rightarrow$ Medium} && 100\% && 0\% \\
        {\em Medium $\rightarrow$ Low} && 100\% && 60\% \\
        {\em Low $\rightarrow$ Low} && 20\% && 0\% \\
        {\em Others}                && 0\%  && 0\% \\ 
        \bottomrule
    \end{tabular}
\end{table}

\verOne { 
We used the dataset \textbf{Data-PerCom} for ambient audio to build the classification model
(F1 of 0.86 and FPR of 9.3\%).  The 100 samples we collected via audio streaming
channels in dataset \textbf{Data-TMC-audio} are fed to the classifier for prediction.
} 
Table~\ref{tbl:audio-streaming} presents the FPR of non co-presence detection
under the streaming attacks over Wi-Fi and cellular data channels. The results
indicate that the attacker (1) has a higher chance of success using the Wi-Fi
channel and (2) could be thwarted when either the ambient audio at \prover is low
frequency or if the ambient audio at \verifier is high frequency.


This simple streaming attack with commodity devices shows that the
audio-only system is highly vulnerable to relay attacks, especially via the
Wi-Fi channel. The attack has very high success rate regardless of hardware
variations and network delays inherent to streaming. However, an attacker can
succeed only when relaying ambient audio from a higher frequency acoustic
environment to a similar or lower frequency acoustic environment, such that,
the higher frequency dominates the lower frequency, and makes \verifier falsely
record \prover's ambient noise instead of the real ``localized'' ambient noise. 

The audio features we used, i.e., the ones proposed in
\cite{DBLP:conf/esorics/HaleviMSX12}, are not sensitive to time
synchronization. This is effective in terms of co-presence detection (i.e.,
results in very low FNR). However, as we can see from our experiments, these
features also enable the attacker to succeed in the relay attack with a very
high chance. Other audio features, such as the ones proposed in
\cite{schurmann2013audio}, require tight synchronization and could be more
resistant to relaying. Unfortunately, because of their high sensitivity to
synchronization, these features did not perform well in the benign
(co-presence) case based on our experiments (i.e., resulted in high FNR).

\subsubsection{Audio-Radio System}
Truong et al. \cite{TruongPerCom14}, evaluated the performance of an
audio-radio system against a unidirectional, \singlemodal attacker. They showed
that while the system achieves good performance (F1 of 0.98) and high security
(FPR of 2.0\%), a contextual attacker could increase the FPR: from 0.18\% to 65.8\%
(manipulating \wifi), from 1.1\% to 1.2\% (\bluetooth); from 1.62\% to 3.01\%
(audio). 
\verOne { 
Now, we will analyze the same system and same dataset \textbf{Data-PerCom} against a bi-directional (for
radio), \multimodal attacker. 
} 
To model the attack, in each run, the non
co-presence samples in the test data were transformed as below.



\noindent\textit{Audio}: Because raw audio data is additive, and one-side
context manipulation for audio is tested, an adversary can be modelled by
replacing \verifier side audio ($X_a$) to be the sum of its own ambient
audio and \prover side audio ($X_a + X_b$).
    
\noindent \textit{Radio} (\bluetooth and \wifi): In \cite{TruongPerCom14}, the
set of radio records from two devices $A$ and $B$ are defined as: $S_{a}
= \{(m^{(a)}_i,s^{(a)}_i)\ |\ i \in \mathbb{Z}_{n_a-1} \}$, and $S_{b}
= \{(m^{(b)}_i,s^{(b)}_i)\ |\ i \in \mathbb{Z}_{n_b-1} \}$, where ($m,s$) with
$m$ is an identifier and $s$ is associated signal strength of a beacon; $n_a$
and $n_b$ denote the number of different beacons (i.e., Wi-Fi access points or
Bluetooth devices). The both-sides contextual adversary can be modeled by
replacing $S_a$ with $S_a \cup \{(m, s) ~\forall (m, s) \in S_b, m \not \in
S^{(m)}_a\}$, and $S_b$ with $S_b \cup \{(m, s) ~\forall (m, s) \in S_a, m \not
\in S^{(m)}_b\}$.


We considered two approaches of fusing sensor data against bi-directional relay
attacks and showed which of them is more suitable for resisting against the
presence of contextual attackers.


\begin{table*}[t] 
\centering 
\caption{
\footnotesize{
\verOne { 
This table shows how well different types of context-manipulating attackers perform in scenarios employing different types of fusion. The horizontal blocks refer to increasingly powerful attackers with the ability to manipulate zero, one or two context sensor modalities.
\textbf{Values in the table are FPRs with/without different contextual attacks in various audio/radio/physical systems}. \textbf{Notations:} Sets of manipulated sensors are put inside curly braces \{\}.  
} 
\{$\widetilde{X}$\} denotes an arbitrary
            set of sensor modalities.  Fuse-F: features-fusion, Fuse-D-S:
            decisions-fusion from single modalities, Fuse-D-M: decisions-fusion
            from subsets of modalities.  \textbf{Result highlights}:
            Manipulation of sensor modalities, especially multiple of them, can
            significantly reduce security (increase FPR) in most cases.
    Decisions-fusion can help improve security when dominant sensors are
    manipulated, but it may reduce usability (increase FNR).
}
\label{tbl:all-mani}}
\resizebox{\textwidth}{!}{%
\begin{tabular}{@{}l c rr c rr c rrr@{}}
\toprule 
&
& 
\multicolumn{2}{c}{{\bf Audio-Radio}}   & \phantom{aaa} 
&
\multicolumn{2}{c}{{\bf Physical}}   &  \phantom{aaa} 
&
\multicolumn{3}{c}{{\bf Audio-Radio-Physical}}  \\
%
\cmidrule{3-4}  \cmidrule{6-7}  \cmidrule{9-11} 
&&  
{\em Fuse-F} &
{{\em Fuse-D-S}} && 
{\em Fuse-F} & 
{\em Fuse-D-S} && 
{\em Fuse-F} & 
{\em Fuse-D-S} & 
{\em Fuse-D-M} \\
%
&& (1) & (2) && (3) & (4) && (5) & (6) & (7) \\
%
\cmidrule{3-4}  \cmidrule{6-7}  \cmidrule{9-11} 
&& 
2.0\%   & 
2.0\%  && 
7.5\%   & 
13.0\%  && 
3.0\%   & 
27.1\%  & 
6.9\%\\                                                          
\multirow{-1}{*}{\begin{rotate}{90} {\bf Zero-modality} \end{rotate}}&& 
\colorbox{gray!90}{(FNR: 1.4\%)}   &
\colorbox{gray!90}{(FNR: 12.0\%)}  && 
\colorbox{gray!90}{(FNR: 3.9\%)}   & 
\colorbox{gray!90}{(FNR: 14.5\%)}  &&
(FNR: 0.0\%)    & 
(FNR: 0.3\%)   & 
(FNR: 0.0\%) \\
&& 
(F1: 0.977)    & 
(F1: 0.925)    && 
(F1: 0.928)    & 
(F1: 0.861)    &&
(F1:0.990)     & 
(F1: 0.923)    & 
(F1: 0.980) \\
\cmidrule{3-4}  \cmidrule{6-7}  \cmidrule{9-11} 
&& 
\{\audio{}\}: 3.0\%       &
\{\audio{}\}: 3.0\%       && 
\{\temperature{}\}: 8.3\%     & 
\{\temperature{}\}: 17.0\%    &&
\{\audio{}\}: 87.7\%     &
\{\audio{}\}: 45.3\%       &
\{\audio{}\}: 36.9\% \\
&& 
\{\bluetooth{}\}: 2.7\%       &
\{\bluetooth{}\}: 9.0\%       && 
\{\gas{}\}: 11.9\%        & 
\{\gas{}\}: 20.0\%        && 
\colorbox{gray!50}{\{\bluetooth{}\}: 100\%}         & 
\colorbox{gray!50}{\{\bluetooth{}\}: 45.8\%} &
\colorbox{gray!50}{\{\bluetooth{}\}: 36.9\%} \\
&& 
\colorbox{gray!20}{\{\wifi{}\}: 99.8\%}       & 
\colorbox{gray!20}{\{\wifi{}\}: 8.0\%}        && 
\{\humidity{}\}: 15.3\%   &
\{\humidity{}\}: 24.4\%   && 
\{\wifi{}\}: 12.3\%    & 
\{\wifi{}\}: 44.8\%   &
\{\wifi{}\}: 35.0\% \\
&&                      
&
&& 
\colorbox{gray!30}{\{\altitude{}\}: 55.1\%}   &
\colorbox{gray!30}{\{\altitude{}\}: 33.1\%}   && 
\{\altitude{}\}: 5.4\%    & 
\{\altitude{}\}: 37.9\%        &
\{\altitude{}\}: 6.9\%\\    
&&                      
& 
&&                      
& 
&& 
\{\gas{}\}: 5.9\%       & 
\{\gas{}\}: 29.6\%      &
\{\gas{}\}: 6.9\%\\    
&&                      
&
&&                      
& 
&& 
\{\humidity{}\}: 3.4\%      &
\{\humidity{}\}: 29.1\%       &
\{\humidity{}\}: 6.9\%\\
\multirow{-3}{*}{\begin{rotate}{90} {\bf Single-modality} \end{rotate}}&&                      
&
&&                      
& 
&& 
\{\temperature{}\}: 3.4\%       &
\{\temperature{}\}: 31.5\%      &
\{\temperature{}\}: 6.9\%\\
%
\cmidrule{3-4}  \cmidrule{6-7}  \cmidrule{9-11}
&& 
\{\audio{}, \bluetooth{}\}: 3.6\%     &
\{\audio{}, \bluetooth{}\}: 96.0\%    && 
\{\gas{}, \temperature{}\}: 13.9\%    & 
\{\gas{},\temperature{}\}: 40.1\%     && 
\colorbox{gray!70}{\{\bluetooth{}\}$\cup$\{$\widetilde{X}$\}: 100\%}             & 
\{2 sensors\}:                      & 
\{\audio{}, \bluetooth{}\}$\cup$\{$\widetilde{X}$\}: \\
&& 
\{\bluetooth{}, \wifi{}\}: 99.8\%     & 
\{\audio{}, \wifi{}\}: 96.0\%         && 
\{\gas{}, \humidity{}\}: 15.7\%       &  
\{\humidity{}, \temperature{}\}: 41.9\% && 
                                      & 
32.0-75.4\%                          & 
$>97.5$\% \\
&& 
\{\audio{}, \wifi{}\}: 100\%          & 
\{\bluetooth{}, \wifi{}\}: 100\%      && 
\{\humidity{}, \temperature{}\}: 29.6\% &  
\{\altitude{}, \temperature{}\}: 50.6\% && 
\{\audio{}\}$\cup$\{$\widetilde{X}$\}$>$74.9\%  & 
\{3 sensors\}:                      &
\{\audio{}, \wifi{}\}$\cup$\{$\widetilde{X}$\}: \\
&& 
\{\audio{}, \bluetooth{}, \wifi{}\}: 100\%      &
\{\audio{}, \bluetooth{}, \wifi{}\}: 100\%      && 
\{\gas{}, \humidity{}, \temperature{}\}: 31.1\%   & 
\{\gas{}, \humidity{}\}: 57.5\%               && 
                                               & 
37.4-97.5\%                                    & 
$>88.2$\% \\
&&                              
&                               
&&
\{\altitude{}\}$\cup$\{$\widetilde{X}$\}: & 
\{\altitude{}, \humidity{}\}: 61.2\%          && 
\{$\widetilde{X}$\}$\backslash$\{\audio{}, \bluetooth{}\}:$<12.3\%$      &
\{4 sensors\}:                              &
\{\altitude{}, \gas{}, \humidity{}, \temperature{}\}:\\

&&                              
&
&& 
64.7-100\%&  
\{\altitude{}, \gas{}\}: 65.5\%               && 
                                       & 
97.5-100\%                                   & 
9.9\% \\

&&                              
&
&& 
&
rest: 100\%     &&                    
                &             
rest: 10\%      & 
\colorbox{gray!70}{\{\bluetooth{}, \wifi{}\}: 36.9\%} \\

\multirow{-4}{*}{\begin{rotate}{90} {\bf Multi-modality}\end{rotate}}&&
&                
&&                             
&                          
&&                    
&                  
&
rest: 6.9-87.7\% \\
\bottomrule
\end{tabular}
}
\end{table*}

Table~\ref{tbl:all-mani} (columns 1 and 2) presents the analysis results of
training model combining all three audio-radio modalities (\audio, \bluetooth
and \wifi) and testing with different attacks. Zero-modality attack shows the
very low FPR with both fusion methods. The FNR for decisions-fusion is higher
compared to that for features-fusion.
For features-fusion, the results are aligned with the ones reported in~\cite{TruongPerCom14}.

In \singlemodal attack, manipulating Wi-Fi, the dominant feature, results in a
very high success rate with features-fusion. The results change when
decisions-fusion was applied.
In such case, manipulating any single sensor, even the most powerful one, does not
significantly degrade the overall security. The FPR in case \wifi was
manipulated decreases from 99.8\% (features-fusion) down to 8\% (decisions-fusion).
We recall that the performance difference of audio and radio sensors is not
large (as reported in \cite{TruongPerCom14}, F1 ranges from 0.857 for \audio to
0.989 for \wifi).
%
This explains why decisions-fusion reduces the overall performance 
slightly (F1 reduces from 0.977 to 0.925) in case of \zeromodal attack but
significantly improves the security under a single-modality attack. The
security is very low in \multimodal attack, and neither of the fusion
approaches could restore the security level when majority of the sensors are
under attacker's control.  
We earlier argued that audio and radio modalities can be manipulated simultaneously.
\subsubsection{Physical System}
\verOne { 
Shrestha et al. \cite{ShresthaFC2014} introduced four physical modalities (\altitude, \humidity, \gas,
and \temperature) for co-presence detection. 
} 
The performance of
the features-fusion based classifier trained with their dataset is good (F1 of
0.957, FPR of 5.81\%) against a \zeromodal adversary.

Based on our attacks against physical modalities (Section \ref{sec:physical}),
we consider an adversarial model where an attacker can manipulate the physical
context on one side (unattended verifier) to match the sensor readings at the
other side (prover). 
\verOne { 
To model this attack, using the dataset \textbf{Data-FC}, we transformed all non co-presence samples in the
test set to the ``attack'' value (distance 0).  
} 
The distance
is set to 0 as data collection in \cite{ShresthaFC2014} was done by a single
device at a given point of time, hence, no hardware effect or calibration error
was taken into account.  The non co-presence class in the dataset is about 18
times larger than co-presence class. To correct this imbalance, we applied the
same under-sampling as in \cite{ShresthaFC2014}: we divided the non co-presence
samples into 19 subsets, ran several rounds of cross validation taking 10
subsets in each round and aggregated the results in the end. In addition to
the features-fusion employed in \cite{ShresthaFC2014}, we tested the 
decisions-fusion similar to our audio-radio system analysis in the previous section.

Table~\ref{tbl:all-mani} (columns 3 and 4) shows our analysis results.  The
system performance in zero-modality attack is well-aligned with the one
reported in \cite{ShresthaFC2014}.  As in \cite{ShresthaFC2014}, among four
physical modalities, \altitude performs the best. Consequently, manipulating
only \altitude degrades the security vastly with features-fusion (FPR increases
to over 50\%). Decisions-fusion in general brings lower security and lower
performance/usability in \zeromodal attack and \singlemodal attack.  However,
it avoids the dominance of sole sensor in case the attacker can control such
sensor (\altitude in this case). Decisions-fusion can also help improve
security against a \multimodal attacker who manipulates \altitude along with
other sensors. Compared to audio-radio system, in physical system, attacking
each single modality results in higher success rate. 
\subsubsection{Audio-Radio-Physical System}

\verOne { 
Unlike the dataset for physical sensors \cite{ShresthaFC2014} which was
collected from one device at a time only, the dataset \textbf{Data-TMC} that we collected for this work contains data from pairs of
devices, and therefore hardware variance and calibration errors between
co-presence device sensors need to be taken into account. 
} 
When we try to model
the contextual attack on given sensor(s), distance 0 does not ensure that the attack will succeed.  As
the classifier is trained with data which may contains noise, we compute the
mode of the histogram for distance values for the co-presence samples. As the data
aggregated is from two participants, histograms of distance values are not
uninomial but multinomial. Multinomial distribution implies several modes. For
each physical sensor, we choose a mode value and assign it as the distance value.
The mode values for \altitude, \gas, \humidity and \temperature are 13.54, 0.3,
6.61 and 0.153, respectively. 
As the manipulation by replacing the radio data at
both sides has to be identical, the distance features for radio sensors are set to 0.

Table~\ref{tbl:all-mani} (columns 5, 6 and 7) reports our analysis results with
different fusion methods. Under \zeromodal attack, features-fusion performs the
best while decisions-fusion from single modalities performs the worst.
Features-fusion uses all possible features for training so that the classifier
can be built based on the best features or best combination of features
(\bluetooth and \audio with our current dataset). Thus, it returns the
best results (in the absence of context manipulation) compared to any other
ways of fusing sensor data. Decisions-fusion based on single modalities lets
the worst sensors being able to contribute to the voting scheme, thus bringing
down the overall performance. This is the case in our dataset where radio
sensors and audio sensor perform better than physical sensors. Note that if all
sensors perform equally well, features-fusion and decisions-fusion would not
differ much. Decisions-fusion from subsets of sensors has a moderate
performance, worse than features-fusion but better than decisions-fusion from
single modalities.  This hybrid approach avoids mis-learning as in the case of
using a single modality only.

Let us now assess the security of this co-presence detection system when any
single modality is controlled by the attacker. Depending on how sensors are
fused, the impact of manipulated sensor varies. In features-fusion, as the
classifier decision relies on the best features of dominant sensors, the FPR
increases drastically when such sensors are manipulated (i.e. \audio or
{\bluetooth} in our dataset). In contrast, when weaker sensors (physical or
{\wifi}) are manipulated, it has a relatively small impact on the security as
the resulting FPR increases a bit compared to a zero-modal attack (especially
for \wifi). Decisions-fusion reduces attacker success rate when single sensor
is manipulated, for example, FPR of manipulating \bluetooth decreases from
100\% (features-fusion) to 36.9\% (decisions-fusion). 
Recall that manipulating single sensor is not difficult as we demonstrated in Section
\ref{sec:manipulation_attack}.

An attacker has the highest chance to succeed if he can control the dominant
sensors or a subset of sensors that contain the dominant sensors. In such case,
the success rate could reach 100\% with only one single dominant sensor (i.e.
\bluetooth in our dataset) if the system uses features-fusion or with
majority dominant sensors (i.e. \audio and \bluetooth). In most cases,
attacking the set of weak sensors (e.g. \{\altitude, \gas, \humidity,
\temperature{}\}) does not impact the security much, except when system uses
decisions-fusion from single modalities.

\section{Discussion, Potential Mitigations and Future Work}
\label{sec:discussion}
\noindent \textbf{Classifier Analysis:}
In our attack analysis, we used classification techniques as implemented by the 
original contextual systems~\cite{ShresthaFC2014,TruongPerCom14}. 
We used different machine learning techniques (Decision Tree and Random Forest) with ten-fold cross
validation implemented in Scikit-learn~\cite{scikit-learn} as mentioned 
in Section~\ref{sec:analysis_method}. When we train/build the classifier model 
using benign training datasets, the classifier chooses the thresholds such that it
yields the best result with high F1 score and low FNRs/FPRs. This model works
best in case there is no context manipulation attack. 
Our results under this setting are inline with that reported 
in original systems~\cite{ShresthaFC2014,TruongPerCom14}. 
For evaluating the performance of our context manipulation attacks, 
we used the exact same classifier models since they demonstrated the best performance in the setting with
no context manipulation attacks.
The classifier model may have performed better if the classification thresholds were modified such that
the model is more robust towards False Positive Rates, i.e., result in lower FPR. 
However, it is important to note that changing the thresholdization setting to
yield low FPR may result in high FNR and thereby decrease the overall F1 score which leads 
to decreasing the usability of the system. Since usability is an important 
attribute of the contextual co-presence detection systems, our analysis
represents a valid attack setting that serves to demonstrate the weaknesses
of, and potential fixes to, the existing usable systems.

\vspace{1mm} 
\noindent \textbf{Reducing Attack Success with Decisions-Fusion:}
In the previous section on analysis of an audio-radio-physical system,  we
showed that decisions-fusion reduces attack success rates in cases where the
minority of the sensors are manipulated.  However, this may come at the cost of
higher FNR which represents the usability of co-presence systems.
%
Decisions-fusion 
from single sensors improves security when individual sensors
perform well. However, it increases the attack success rate for weak sensors as
they equally contribute to the voting. For example, in the context of the audio-radio-physical
system, attacking weak sensors such as \humidity or \gas brings relatively high
success rate compared to features-fusion. 
Decisions-fusion
from subsets of sensors reduces the FPR in general especially when
dominant sensors are controlled by the attacker

\vspace{1mm} \noindent \textbf{Other Potential Countermeasures:}
%
%
%
%
Typically, during the authentication/deauthentication process, the prover moves
nearer to/farther away from the verifier. In this case, the radio signals
changes gradually, i.e., if prover and verifier move towards APs, then new APs
will be shown, or their signal strengths will continuously grow, while if they
move further away from APs, their strengths will decrease or the APs will not
be visible at all. If the verifier or prover device detects much more APs (or
Bluetooth devices) nearby all of a sudden, it probably indicates a radio
manipulation attack. The system can be made aware of such situations.

We noticed that when the verifier is in an environment which has high frequency
noise, an attacker tends to fail with audio streaming. This can be used to
design an active defense mechanism such that whenever audio contextual
information is requested, the verifier can emit a high frequency, potentially random audio sounds. This
audio signal can be for a short duration, and does not need to be loud (not
high amplitude). As a result, the chances of attacker succeeding in a
relay attack could be reduced.

When an authentication request has been initiated or finished, the user can be
passively notified at both devices.  Passive notification can be a flashing of
LED light or  beep on the prover device (key or phone).  Hence, even if the
verifier device is left unattended, user may notice at the end of the prover device that
someone is trying to authenticate to the verifier, or has authenticated on user's
behalf. Whether or not users would actually pay attention to such notifications
should be subject to further scrutiny. It may help reduce the risk of
context-manipulation relay attacks.


\vspace{1mm}
\noindent \textbf{Limitations:}
There are certain limitations of our work.
\verTwo{Our attack experiments were done in lab settings under normal 
environmental conditions. It may require more effort or sophistication from the adversary in real-life settings,
under an arduous environment, to make these attacks work. For example, using ice cubes to 
decrease temperature below freezing point may not work, or using a hair dryer in a very 
cold outdoor environment may make it difficult to increase the temperature to a desired value, 
accessing a sensor when the sensor placement is unknown or access to sensor is difficult, and so on.
Nevertheless, our work demonstrates that the designers of contextual security systems should consider such attacks 
seriously while developing such solutions.}
The dataset we used for analyzing the attacks in audio-radio-physical system is
relatively small. It was collected from limited number of devices. It might not
represent all possible scenarios and environments.  However, it
was sufficient to demonstrate the impact of attacks and defensive solutions. It
gave insights for better understanding of the contextual co-presence
detection system and possible defenses to improve security against different
contextual attacks.
Further work may be needed to collect and analyze a
larger scale dataset to evaluate this system.  The decisions-fusion from
subsets of sensors seems to be the most appropriate solution for improving
security against context manipulation attacks. However, we have analyzed it
only with three subsets: acoustic (\audio), radio ({\bluetooth, \wifi}) and
physical subsets ({\altitude, \gas, \humidity, \temperature}). In design of a
real system in the future, we would like to test different subsets combinations
to find the best candidate for fusion.


\section{Conclusions} \label{sec:conclude}
Contextual co-presence detection has been shown to be a very promising relay
attack defense in many mobile authentication settings suitable for
off-the-shelf, sensor-equipped devices.  We presented a systematic assessment
of co-presence detection in the presence of a context-manipulating attacker. 
Our work suggests that tampering with the context can be achieved with simple
yet effective strategies, and the security offered by co-presence detection is
therefore weaker than previously believed. 
We also suggested potential countermeasures (e.g., decisions-fusion based
machine learning \verOne{classification technique},
that may be used to strengthen the security of co-presence
detection against a \multimodal attacker. Some of these countermeasures may
require a thorough future investigation, which we plan to pursue.  
%

\section*{Acknowledgments} \label{sec:ack}
The authors would like to thank TMC reviewers for their useful feedback on the paper. 
This work was supported in part by NSF grants (CNS-1526524 and CNS-1547350), 
the Academy of Finland (projects 274951 and 309994) and Intel (ICRS-SC).

\bibliographystyle{abbrv}
\bibliography{main,all-fc}
\newpage

\appendices \label{sec:appendix}

\section{Increasing the temperature when the attacker does not know $VS$'s location} 
\label{sec:a1}
\begin{figure}[htpb]
    \begin{minipage}[b]{\linewidth}
      \centering
      \includegraphics[width=\textwidth]{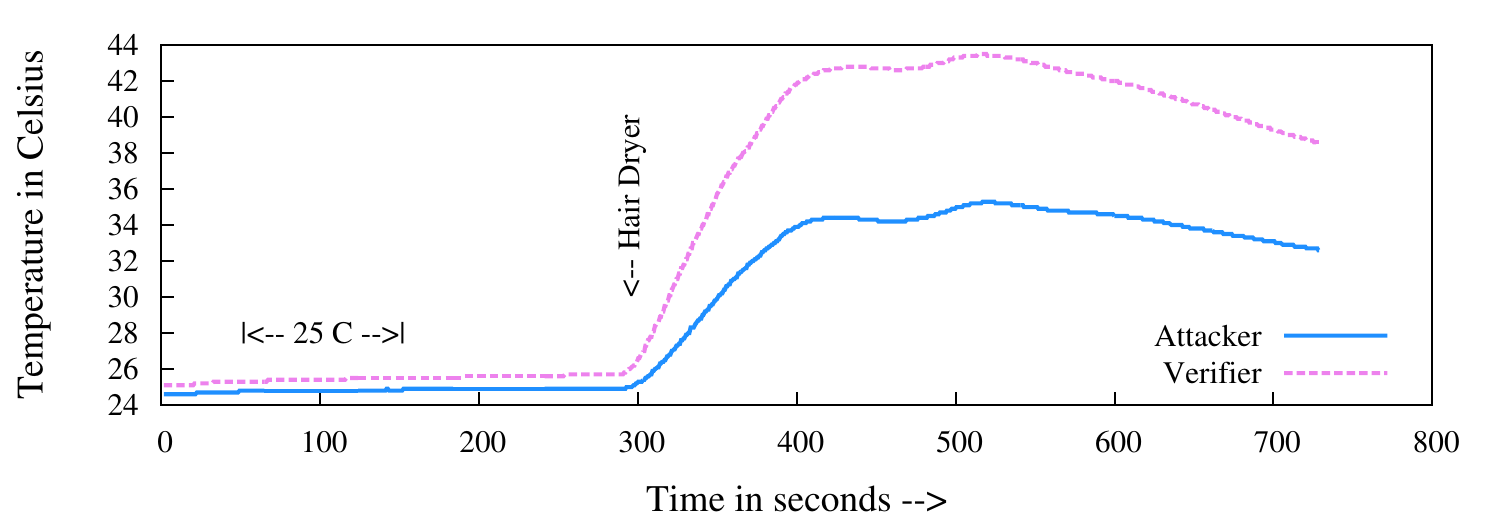}
      \caption{Increasing the temperature; location of $VS$ unknown to the attacker; $VS$ is 
      10 cm closer to hair dryer than $FS$; the attacker trying to increase temperature to 35 
      \textdegree C.}
      \label{fig:HD_AVH}
    \end{minipage}%
\end{figure}

\begin{figure}[htpb]
    \begin{minipage}[b]{\linewidth}
      \centering
      \includegraphics[width=\textwidth]{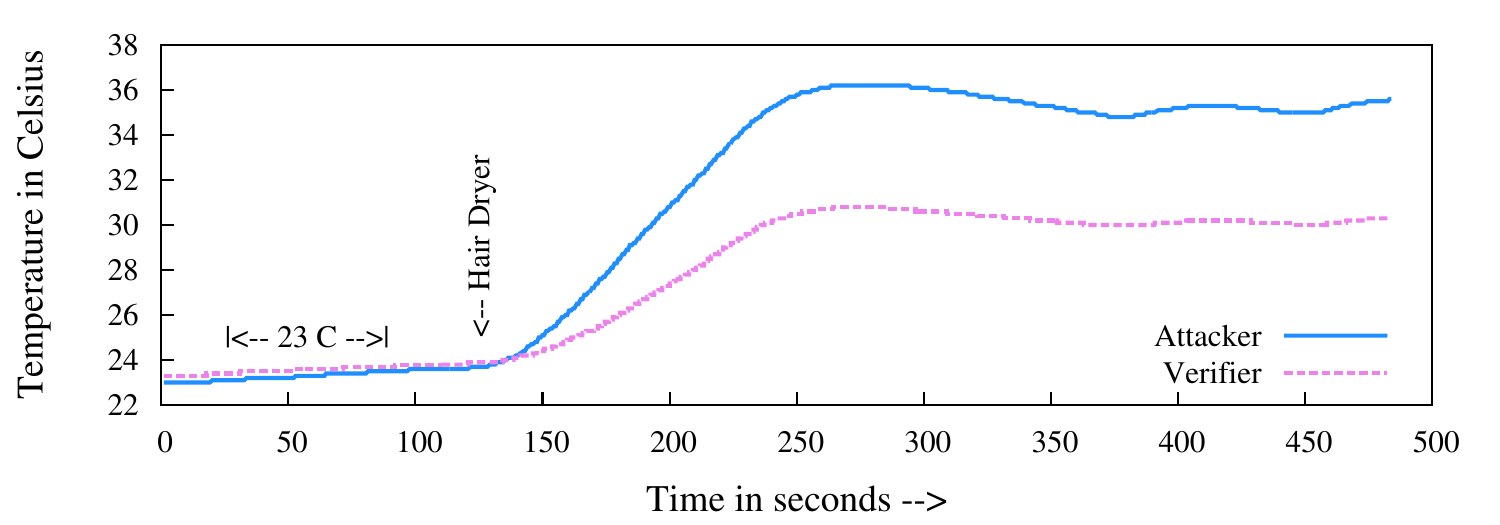}
      \caption{Increasing the temperature; location of $VS$ unknown to the attacker; $FS$ is 
      10 cm closer to hair dryer than $VS$; the attacker trying to increase temperature to 35 
      \textdegree C.}
      \label{fig:HD_VAH}
    \end{minipage}
\end{figure}

An attacker who \verTwo{does not} know the location of $VS$ will try to keep the $FS$ as 
close as possible and perform the attack activity. 
We placed the $FS$ 10 cm apart from the $VS$ and performed 
experiment in two settings.
In the first setting, the hair dryer is closer to $VS$ as shown 
in Fig. \ref{fig:HD_AVH}, and in the latter setting, 
the hair dryer is closer to $FS$ as shown in Fig. \ref{fig:HD_VAH}.


\vspace{2mm}
\section{Increasing the CO gas level}
\label{sec:a2}
\begin{figure}[htpb]
    \begin{minipage}[b]{\linewidth}
      \centering
      \includegraphics[width=\textwidth]{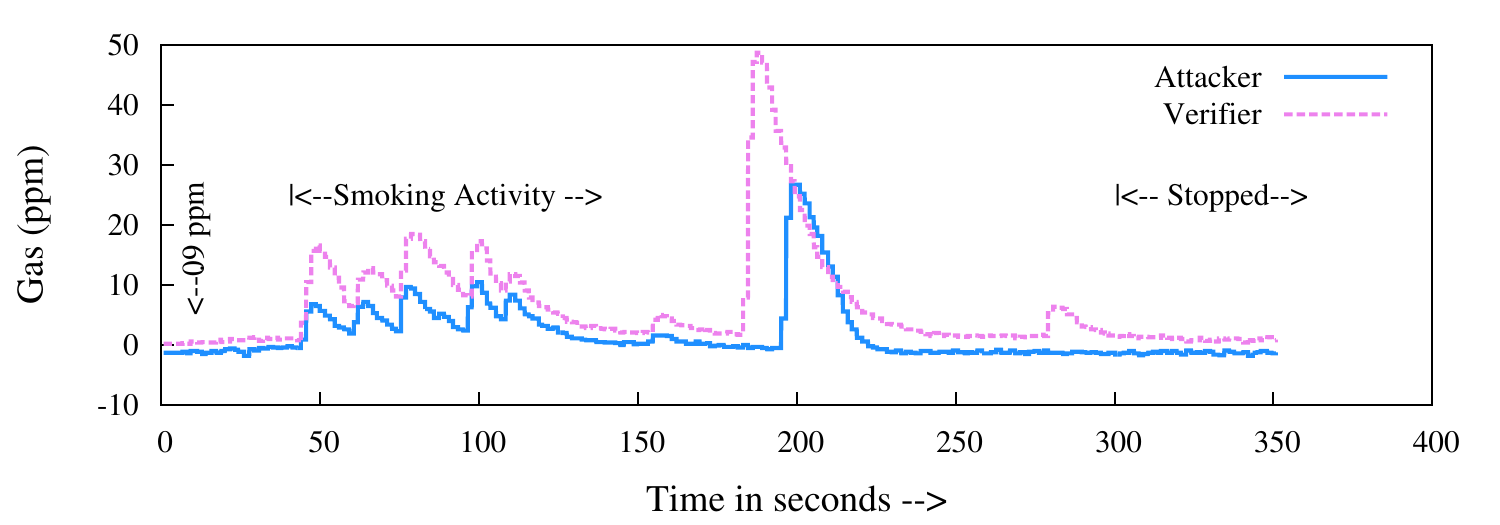}
      \caption{Effect of cigarette in CO level; increasing the gas content to an arbitrary value and waiting to decrease to desired level.}
      \label{fig:gas_cigarette}
    \end{minipage}
\end{figure}

\begin{figure}[htpb]
    \begin{minipage}[b]{\linewidth}
      \centering
      \includegraphics[width=\textwidth]{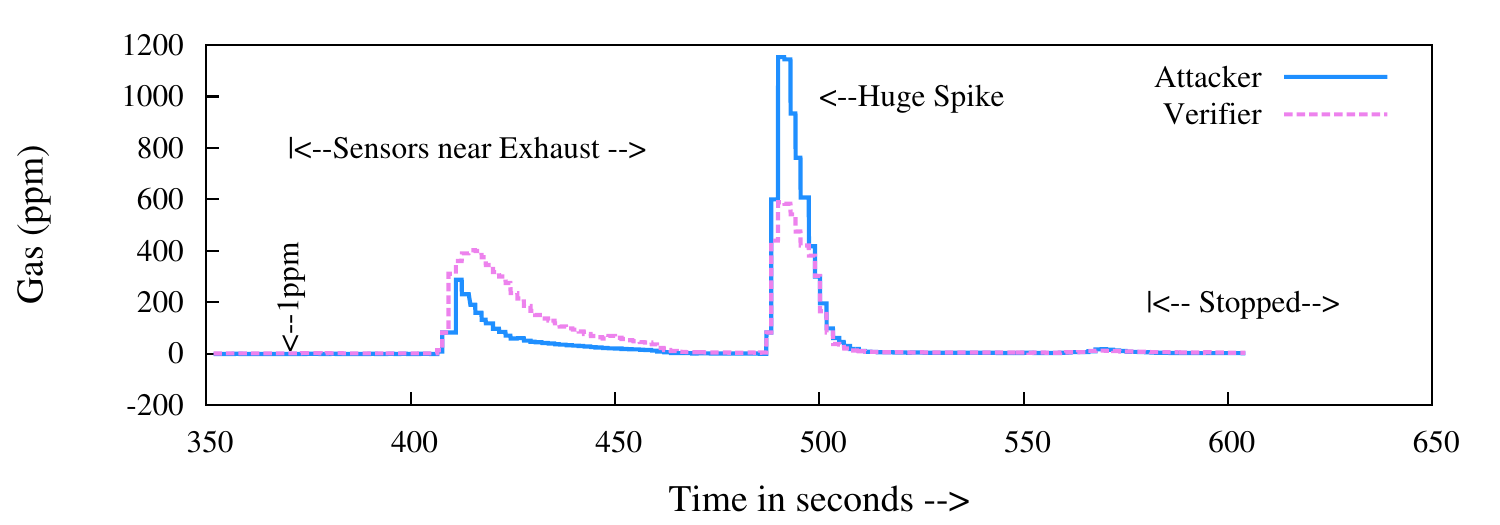}
      \caption{Effect of car exhaust in CO level; increasing the CO gas level to arbitrary 
      value and wait to decrease to desired level.}
      \label{fig:gas_carExhaust}
    \end{minipage}%
\end{figure}

We effectively manipulated the CO gas sensor using cigarette and car exhaust. The increase 
in the gas level due to the activity is abrupt when CO is blown onto the sensors, however, it takes 
a while for the sensors to fall back to normal readings. This provides an enough time window for the attacker as depicted in Figs \ref{fig:gas_cigarette} and~\ref{fig:gas_carExhaust}.


\vspace{2mm}
\section{Increasing the altitude using a car vacuum} 
\label{sec:a3}
\begin{figure}[htpb]
    \begin{minipage}[b]{\linewidth}
      \centering
      \includegraphics[width=\textwidth]{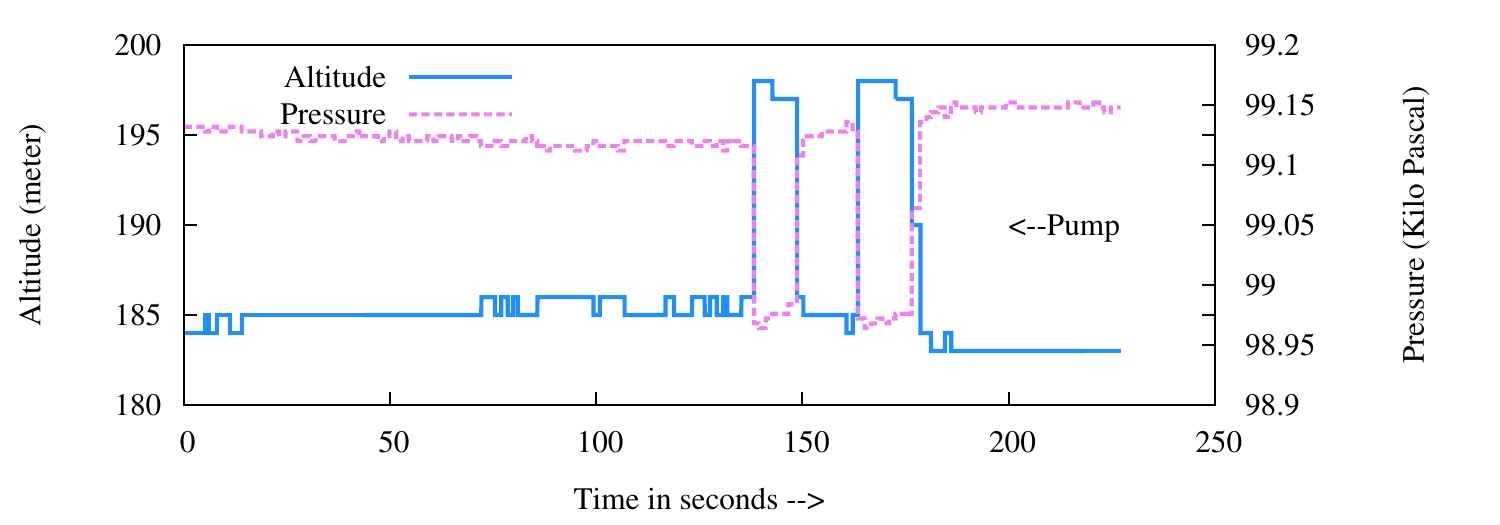}
      \caption{Using a car vacuum cleaner to reduce pressure around the sensor and increase the altitude.}
      \label{fig:carVacuum}
    \end{minipage}%
\end{figure}

As an alternative to air pump, we tried a portable car vacuum cleaner for
inducing an altitude increase. When we hovered the vacuum cleaner pipe around
the sensors, it did not have any effect. However, when we put the pipe just on
top of the sensor, it increased the altitude by 10-11 meters as shown in Fig.
\ref{fig:carVacuum}. An attacker can adjust the altitude to a desired level by
changing the power level of the vacuum cleaner, similar to the air pump
manipulation. The earlier part of the Fig. \ref{fig:carVacuum} shows a little
fluctuation in altitude when we hovered the pipe around the sensors while the
later spikes clearly show that there was an increase of almost 10 meters when
the pipe was touched to the sensors.  A video demo of our attack has been
uploaded to YouTube \cite{carVacuum} to show the effect of portable car vacuum
cleaner on the pressure/altitude sensors.

\begin{IEEEbiography}[{{\includegraphics[width=25mm,height=32mm,clip,keepaspectratio]{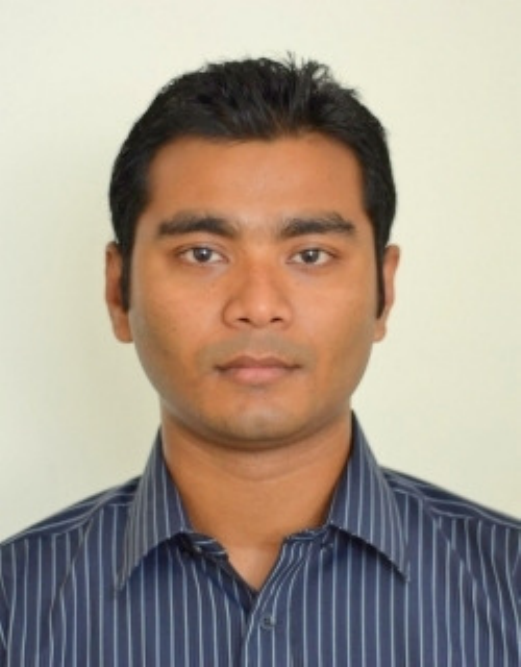}}}]
{Babins Shrestha} is a Senior Information Security Analyst at VISA Inc., Austin, TX. He works in
Cyber Defense team at VISA where he facilitates the detection and prevention of advanced emerging
cyber threats through the development of new tools, techniques, automation, and sensor enrichment.
Shrestha received his Ph.D degree in Computer and Information Sciences from University of Alabama at 
Birmingham (UAB) in 2016. He worked with Dr. Nitesh Saxena as a member of SPIES(Security and Privacy In 
Emerging computing and networking Systems) lab at UAB. He has several journal and conference papers 
published corresponding to his work. He also has four years of work experience in Web Application 
Development at Verisk Information Technologies, previously known as D2Hawkeye Services where he worked 
as Senior Software Engineer. 
\end{IEEEbiography}

\begin{IEEEbiography}[{{\includegraphics[width=25mm,height=32mm,clip,keepaspectratio]{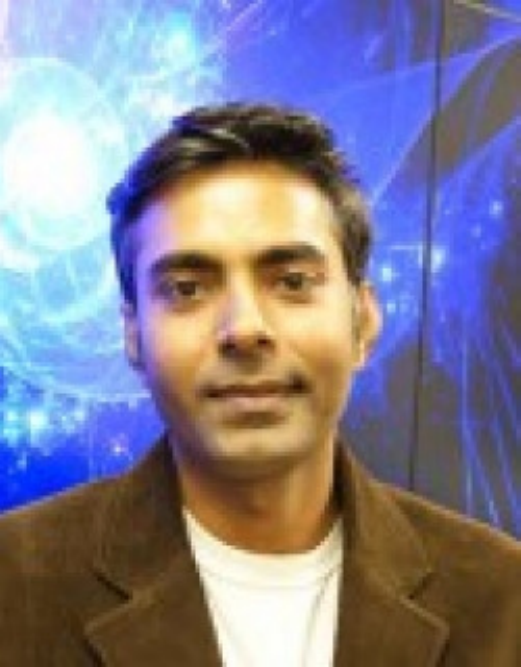}}}]
{Nitesh Saxena} is an Associate Professor of Computer and Information Sciences at the University of 
Alabama at Birmingham (UAB), and the founding director of the Security and Privacy in Emerging Systems 
(SPIES) group/lab. He works in the broad areas of computer and network security, and applied 
cryptography, with a keen interest in wireless and mobile device security, and the emerging field of 
usable security. Saxena's current research has been externally supported by multiple grants from NSF 
and NIJ, and by gifts/awards/donations from the industry, including Google (2 Google Faculty Research 
awards), Cisco, Comcast, Intel, Nokia and Research in Motion. He has published over 110 journal, 
conference and workshop papers, many at top-tier venues in Computer Science, including: IEEE 
Transactions, ISOC NDSS, ACM CCS, ACM WWW, ACM WiSec, ACM ACSAC, ACM CHI, ACM Ubicomp, IEEE Percom, 
IEEE ICME and IEEE S\&P. On the educational/service front, Saxena currently serves as the director and 
principal investigator for the UAB's Scholarship for Service (SFS) program and a co-director for UAB's 
MS program in Computer Forensics and Security Management. He serves as an Associate Editor for flagship 
security journals, IEEE Transactions on Information Forensics and Security (TIFS), and Springer's 
International Journal of Information Security (IJIS). Saxena's work has received extensive media 
coverage, for example, at NBC, MSN, Fox, Discovery, ABC, Bloomberg, MIT Tech Review, ZDNet, ACM 
TechNews, Yahoo! Finance, Communications of ACM, Yahoo News, CNBC, Slashdot, Computer World, Science 
Daily and Motherboard.
\end{IEEEbiography}

\begin{IEEEbiography}[{{\includegraphics[width=25mm,height=32mm,clip,keepaspectratio]{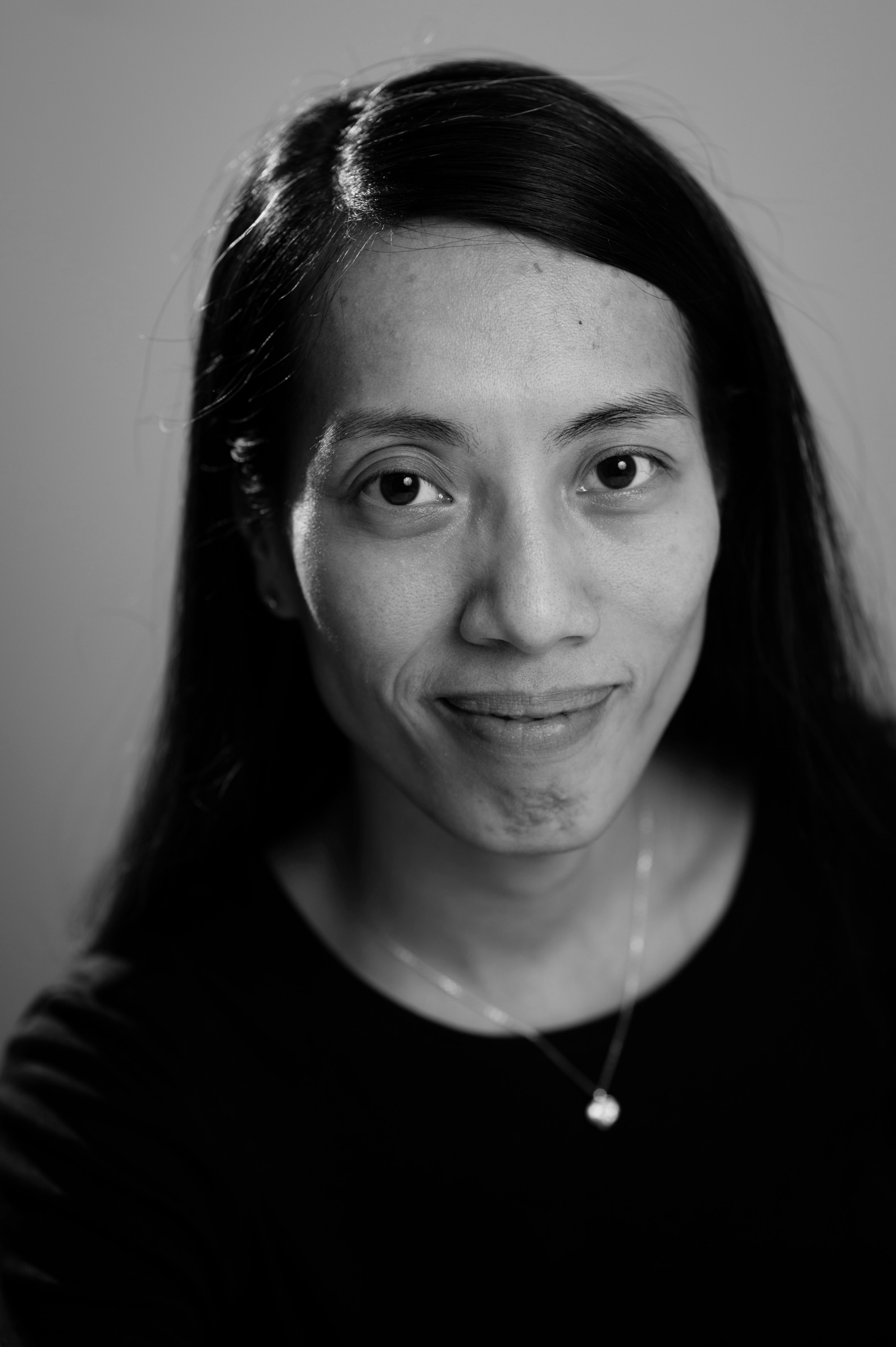}}}]
{Hien Truong} received her Ph.D. degree in Computer Science from INRIA and Universit\'e de Lorraine, 
France, in 2012. From 2013 to 2016, she conducted her postdoctoral research at the University of 
Helsinki, Finland. Since 2016, she is a Research Scientist in security at NEC Laboratories Europe in 
Germany. She has over 10 years of R\&D experience gained while working in Vietnam, Japan, France, 
Finland and Germany. Her research interests cover the design and analysis of system security; privacy 
and trust for distributed systems; and mobile security. Recently, she focuses on applying machine 
learning and data analysis techniques to solve usable security problems.
\end{IEEEbiography}

\begin{IEEEbiography}[{{\includegraphics[width=25mm,height=32mm,clip,keepaspectratio]{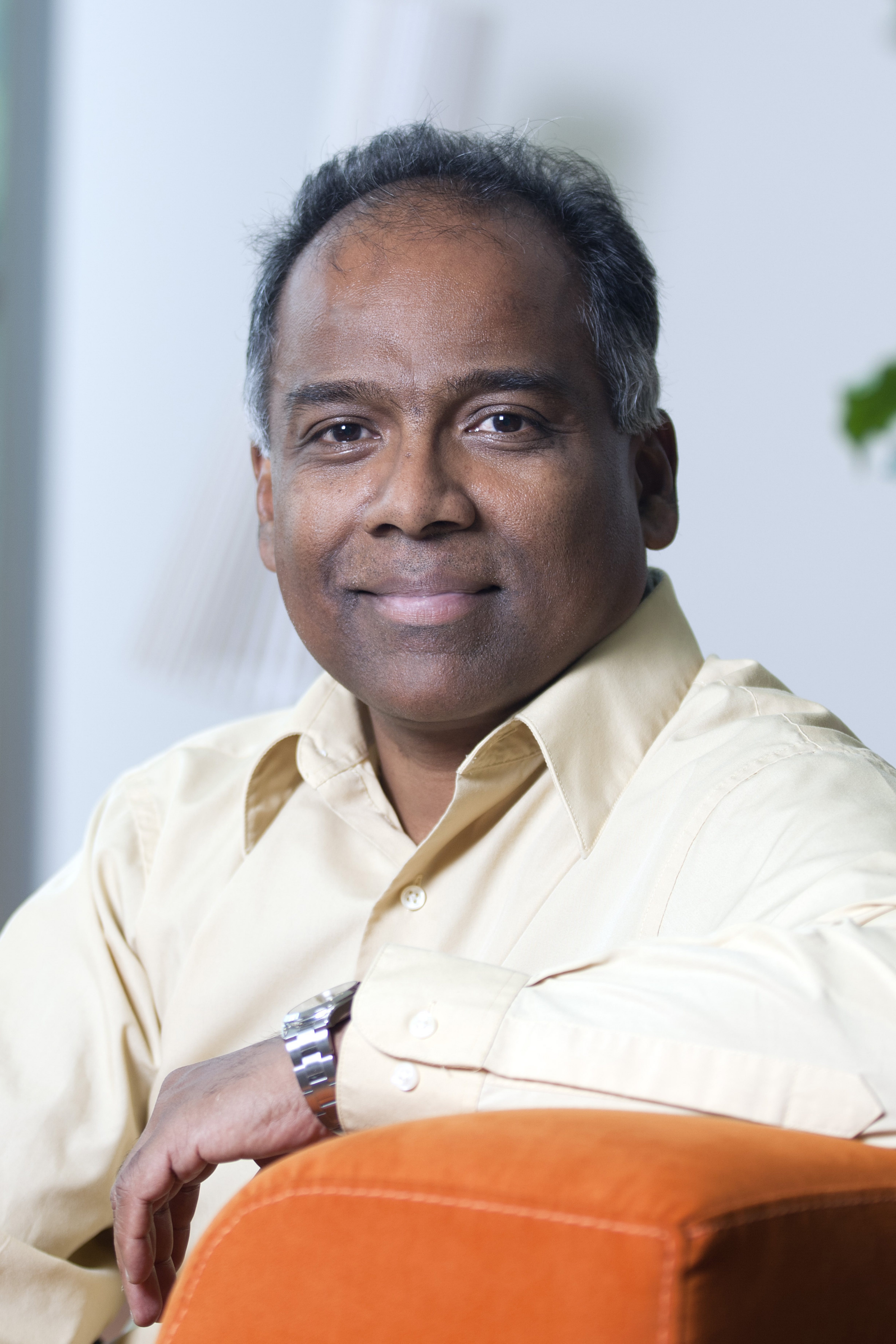}}}]
{N. Asokan} is a Professor of Computer Science at Aalto University where he directs Helsinki-Aalto 
Center for Information Security -- HAIC. He is a Fellow of IEEE and an ACM Distinguished Scientist.
\end{IEEEbiography}

\end{document}